\documentclass[lettersize,journal]{IEEEtran}
\usepackage{amsmath,amssymb,amsfonts}
\usepackage{array}
\usepackage{textcomp}
\usepackage{stfloats}
\usepackage{url}
\usepackage{verbatim}
\usepackage{graphicx}
\usepackage{cite}
\usepackage{siunitx}
\usepackage{bm}
\usepackage{soul}
\usepackage{xurl}
\usepackage{subcaption}
\usepackage{enumitem}
\usepackage{color}
\usepackage{sidecap, caption}

\newcommand{\Deleted}[1]{}

\usepackage{multirow}%
\usepackage{amsthm}%
\usepackage{mathrsfs}%
\usepackage{xcolor}%
\usepackage{textcomp}%
\usepackage{manyfoot}%
\usepackage{booktabs}%
\usepackage{algpseudocode}%
\usepackage{listings}%
\usepackage{eurosym}
\usepackage{acronym}
\usepackage{hyperref}

\newcommand{\citep}[1]{{\cite{#1}}}

\newif\ifincludesupplemental
\includesupplementaltrue 
\newcommand{\suppalt}[2]{\ifincludesupplemental#1\else#2\fi}

\newif\ifincludemaintext
\includemaintexttrue

\acrodef{VR}[VR]{\emph{Virtual Reality}}
\acrodef{AR}[AR]{\emph{Augmented Reality}}
\acrodef{XR}[XR]{\emph{Extended Reality}}
\acrodef{EDA}[EDA]{\emph{Electrodermal Activity}}
\acrodef{HR}[HR]{\emph{Heart Rate}}
\acrodef{HMD}[HMD]{\emph{Head-Mounted Display}}
\acrodef{HRV}[HRV]{\emph{Heart Rate Variability}}
\acrodef{HPA}[HPA]{\emph{Hypothalamic-Pituitary-Adrenal}}
\acrodef{SSQ}[SSQ]{\emph{Simulator Sickness Questionnaire}}
\acrodef{FMS}[FMS]{\emph{Fast Motion Sickness Scale}}
\acrodef{EEG}[EEG]{\emph{Electroencephalography}}
\acrodef{ECG}[ECG]{\emph{Electrocardiography}}
\acrodef{EMG}[EMG]{\emph{Electromyography}}
\acrodef{EOG}[EOG]{\emph{Electrooculography}}
\acrodef{EGG}[EGG]{\emph{Electrogastrography}}
\acrodef{VIMSSQ}[VIMSSQ]{\emph{Visually Induced Motion Sickness Susceptibility Questionnaire}}
\acrodef{RT}[RT]{\emph{Reaction Time}}
\acrodef{CV}[CV]{\emph{Coefficient of Variation}}
\acrodef{VIMS}[VIMS]{\emph{Visually Induced Motion Sickness}}
\acrodef{IPC}[IPC]{\emph{In Person Cohort}}
\acrodef{VMC}[VMC]{\emph{VERA Mirror Cohort}}
\acrodef{VFC}[VFC]{\emph{VERA Full Cohort}}

\begin{document}

\title{Beyond the Lab: Large-Scale Remote Cybersickness Research in Virtual Reality Using the VERA Platform}

\author{Matt Gottsacker, Gerd Bruder, Daniel Zielasko, Alexander Giovannelli, Ali Haskins, Corey Clements,\\ Chloe Beato, John Murray, Robert W. Lindeman, Rui Xie, Jonathan Beever, Nicholas Alvaro Coles,\\ Valerie Jones Taylor, Tabitha Peck, Jeremy Bailenson, and Greg Welch
\thanks{Matt Gottsacker, Gerd Bruder, Alexander Giovannelli, Ali Haskins, Corey Clements, Chloe Beato, John Murray, Rui Xie, Jonathan Beever, and Greg Welch are with the University of Central Florida.
\newline E-mail: \{mattg, bruder, gio, ahaskins, corey.clements, chloe.beato, jtm, rui.xie, jonathan.beever, welch\}@ucf.edu.}
\thanks{Daniel Zielasko is with the Technical University of Denmark.
\newline E-mail: danzi@dtu.dk}
\thanks{Robert W. Lindeman is with the University of Canterbury, Christchurch.
\newline E-mail: rob.lindeman@canterbury.ac.nz}
\thanks{Nicholas Alvaro Coles is with the University of Florida.
\newline E-mail: ncoles@ufl.edu}
\thanks{Valerie Jones Taylor is with Rutgers University.
\newline E-mail: vj.taylor@rutgers.edu}
\thanks{Tabitha Peck is with Davidson College.
\newline E-mail: tapeck@davidson.edu}
\thanks{Jeremy Bailenson is with Stanford University.
\newline E-mail: bailenso@stanford.edu}
\thanks{\newline
This work has been submitted to the IEEE for possible publication. Copyright may be transferred without notice, after which this version may no longer be accessible.}
}

\markboth{Transactions on Visualization and Computer Graphics,~Vol.~X, No.~X, August~2026}%
{Gottsacker \MakeLowercase{\textit{et al.}}: Beyond the Lab: Large-Scale Remote Cybersickness Research in Virtual Reality Using the VERA Platform}

\maketitle

\begin{abstract}
Cybersickness remains a major barrier for the adoption of virtual reality (VR), yet most existing knowledge is derived from laboratory-based studies with relatively small and homogeneous participant samples.
In this paper, we investigate whether remote VR studies can produce cybersickness findings comparable to traditional in-lab experiments while enabling larger and more diverse participant populations.
Using the Virtual Experience Research Accelerator (VERA), we deployed a remote adaptation of the standardized Cybersicker testbed for cybersickness research and collected data from N=263 participants using their own consumer VR headsets.
We compared these results against a previously published in-lab dataset and a demographically matched subset of the remote sample.
Across cohorts, cybersickness outcomes were consistent in direction and temporal pattern, including symptom onset trajectories, Fast Motion Sickness Scale (FMS) ratings, and Simulator Sickness Questionnaire (SSQ) responses, supporting the validity of remote cybersickness human-subjects research.
Leveraging the larger remote dataset, we additionally examined demographic and individual-difference factors associated with cybersickness.
These analyses confirm at scale the effects of sex, sickness susceptibility, and sickness expectation reported in smaller laboratory samples, and they add well-powered evidence on the contested relationship between age and cybersickness.
These findings demonstrate that the VERA platform provides a viable means of remotely replicating laboratory-based studies and highlight its ability to support large, demographically diverse participant samples.
This work establishes a validated methodological foundation for future large-scale remote studies of cybersickness and other VR research topics.
\end{abstract}


\newpage
\section{Introduction}
\label{sec_introduction}

\IEEEPARstart{A}{} primary challenge in the field of virtual reality (VR) is \emph{cybersickness} \cite{Kemeny2020}. 
Characterized by symptoms such as nausea, disorientation, dizziness, and visual discomfort \cite{Gavgani2017}, cybersickness can degrade user experience and limit the duration and effectiveness of VR applications across domains including training, education, and entertainment.
Despite decades of research \cite{Ang2023}, the underlying causes of cybersickness are still not fully understood, and its manifestation varies substantially across individuals, applications, and system configurations.

A key limitation of prior research is that much of our current understanding of cybersickness is derived from laboratory-based studies, which are conducted with relatively small and homogeneous participant populations \cite{Mottelson2017}.
These studies often rely on convenience sampling, frequently recruiting university students or local participants, which introduces biases related to age, education level, geography, cultural background, and prior VR exposure \cite{Saffo2021}.
As a result, findings lack generalizability to the broader population of VR users, particularly as VR technologies become increasingly accessible to diverse audiences. 
Furthermore, the logistical constraints of in-person studies, such as limited throughput, required physical lab space, and participant scheduling, make it challenging to collect large-scale datasets necessary for understanding individual differences and complex interaction effects \cite{Rupert2017}.

Recent advances in remote experimentation infrastructure show promise, as they do not suffer from the same limitations \cite{Ratcliffe2021}. 
Platforms such as the Virtual Experience Research Accelerator (VERA) enable researchers to deploy VR studies outside a VR laboratory, allowing participants to take part in virtual experiences in their own environments using consumer-grade head-mounted displays (HMDs).
This shift presents an important opportunity to scale data collection, increase demographic diversity, and capture more ecologically valid usage conditions. 
However, it also raises a critical research question: \emph{To what extent are results obtained from remote VR studies comparable to or different from those collected under controlled in-lab conditions?}
Establishing this comparability is essential before remote methodologies can be confidently adopted for widespread empirical research in the field of VR.

In this paper, we address this question through a large-scale study on cybersickness that bridges controlled in-lab and remote deployment. 
Building on a well-established in-lab cybersickness testbed \cite{zielasko2024cybersicker}, we adapt the experimental protocol for a remote study using VERA, enabling data collection from a substantially larger and more diverse participant population. 
Our study preserves the core characteristics of the original controlled setup while introducing the flexibility and scalability of remote participation.

We make two primary contributions.
First, we demonstrate that key findings obtained through the remote study design closely align with those reported in a corresponding in-lab experiment, providing empirical support for the validity of remote VR studies in the context of cybersickness research.
Second, leveraging a larger and more diverse sample, we present additional analyses of demographic and individual-difference factors, which confirm and consolidate at scale effects previously reported in smaller laboratory studies and add well-powered evidence on the effect of age.
These contributions establish a firm empirical basis for remote VR studies to complement and extend established in-lab methodologies.
Future studies can build on our methodology to examine other VR research questions that benefit from high levels of demographic control and inclusivity, scale, and ecological validity.

The remainder of this paper is structured as follows.
Section~\ref{sec_rw} reviews background literature on cybersickness, experimental testbeds, and in-person versus remote study methodologies.
Section~\ref{sec_vera} provides an overview of VERA, the platform used to collect data from remote participants.
Section~\ref{sec_method} describes our study design, apparatus, and procedures.
Section~\ref{sec_results} presents the results, including comparisons to prior in-lab findings and analyses enabled by our expanded dataset.
Section~\ref{sec_discussion} discusses the implications of our findings, and Section~\ref{sec_conclusion} concludes with directions for future work.

\section{Background}
\label{sec_rw}

\subsection{Cybersickness}
Cybersickness has been a longstanding issue in the use of immersive technologies. 
While it shares symptoms with motion and simulator sickness, including dizziness, stomach awareness, and eyestrain~\cite{Kennedy1992}, cybersickness is often characterized by greater symptom severity~\cite{Stanney1997-CSnotSS}. 
Symptom presentation can vary substantially across individuals~\cite{Rebenitsch2016}, making the underlying mechanisms of cybersickness difficult to determine. 
As a result, several explanations for cybersickness have emerged, including three widely cited theories: (1) sensory conflict theory, which attributes cybersickness to mismatches among sensory signals related to body motion and orientation~\cite{Reason1975motion}; (2) postural instability theory, which links cybersickness to sustained instability in postural control~\cite{Riccio1991}; and (3) poison theory, which suggests that sensory conflicts trigger a protective physiological response because the body interprets them as signs of poisoning~\cite{Treisman1977}.

In addition to these theories, prior work has identified numerous factors associated with cybersickness related to hardware characteristics, user demographics, and application design~\cite{Chang2020}. 
Hardware-related factors include positional tracking errors~\cite{Biocca1992}, latency between user actions and system responses~\cite{Pausch1992, Caserman2019}, and refresh rate fluctuations~\cite{Harwood1987}. 
Cybersickness susceptibility has also been shown to vary across user demographics and physiological conditions, including sex~\cite{Park2006, Caserman2021, Kelly2023gender}, age~\cite{Paillard2013, Saredakis2020factors, Dilanchian2021}, and health~\cite{LaViola2000, Kelly2026}.
Additionally, application-specific factors such as exposure duration~\cite{Melo2018}, use case~\cite{Davis2014}, and degree of control over navigation~\cite{Giovannelli2025} can influence cybersickness severity.

Given the impact of cybersickness on the use of immersive technologies, researchers have investigated numerous objective and subjective methods for assessing it. 
Common objective measures include biophysical signals such as electrocardiogram (ECG), electromyography (EMG), electroencephalography (EEG), electrodermal activity (EDA), heart rate, and eye movement, all of which have been correlated with cybersickness~\cite{Kiryu2007, Munafo2017TheVR, Dennison2016, Garcia2019development, Liao2020, Islam2022towads, Zielasko2026}. 
Tracking data has also been explored as an objective assessment method, with cybersickness associated with changes in postural sway and movement velocity~\cite{SoHoLo2001, DeGuzman2025}.
Despite the availability of objective assessment methods, questionnaires remain among the most widely adopted approaches for measuring cybersickness, as symptom severity is inherently subjective and is often best captured through self-report~\cite{Davis2014}.
Commonly used self-report instruments include the Simulator Sickness Questionnaire (SSQ)~\cite{Kennedy1993}, the Fast Motion Sickness Scale (FMS)~\cite{Keshavarz2011validating}, which is frequently administered using an adapted 0--10 rating scale~\cite{zielasko2018dynamic}, the Motion Sickness Susceptibility Questionnaire (MSSQ)~\cite{golding1998motion}, and the Visually Induced Motion Sickness Susceptibility Questionnaire (VIMSSQ)~\cite{Keshavarz2023vimssq}.

Despite this research attention, the severity and manifestation of cybersickness still vary considerably across individuals~\cite{Zielasko_Subject2021}.
Large-scale cybersickness data from diverse participant populations is therefore important for improving the generalizability of findings and characterizing these individual differences.

\subsection{Cybersickness Testbeds}
\label{sec_rw_testbeds}
In recent years, researchers have developed several frameworks designed to induce controlled amounts of cybersickness in order to better understand its onset and the factors influencing its severity during immersive experiences.
For instance, Tian et al.\ proposed the Cybersickness Assessment Framework, which examined influential categories related to the environment, experiment design, locomotion, and user vision~\cite{Tian2025cybersickness}.
The framework provided core utilities across these categories to support cybersickness research while also recommending standardized data collection practices.
Using stationary, on-rails, and user-controlled locomotion scenarios, Calandra and Lamberti evaluated cybersickness induced through body-centric and vehicle-centric locomotion~\cite{Calandra2024}.
Their testbed was used to assess the effectiveness of cybersickness mitigation techniques across different movement conditions.
To investigate mitigation techniques triggered at the onset of cybersickness symptoms, Ahmed et al.\ introduced a testbed incorporating deep learning-based symptom detection~\cite{Ahmed2025}.
Their study focused on adversarial attacks capable of preventing symptom mitigation by ``fooling'' the detection models.
Extending cybersickness research beyond single-user settings, Fieffer et al.\ introduced an open-source multiplayer platform centered around virtual maze exploration~\cite{Fieffer2025}.
Their testbed enabled experimenters to customize world design and locomotion factors to induce cybersickness while users completed interdependent collaborative tasks.
Collectively, these platforms demonstrate growing interest in reusable and configurable infrastructures for cybersickness research.

For our work, we aimed to investigate single-user cybersickness experiences under controlled self-motion conditions that induce illusions of translational and rotational movement without the integration of mitigation techniques.
To support this investigation, we leveraged the open-source VR sickness testbed called ``Cybersicker''~\cite{zielasko2024cybersicker}.
Cybersicker was designed to provide a realistic experience modeled after a real amusement park ride despite its role as a controlled cybersickness-inducing simulator (see Figure~\ref{fig_cybersicker}a).

Since its release, the platform has been adopted in multiple VR studies investigating cybersickness and user experience.
For instance, Plaza et al.\ used Cybersicker to design and evaluate interfaces for real-time self-reporting of cybersickness symptom intensity~\cite{Plaza2026}.
Their findings suggested that users generally preferred continuous self-report interfaces over discrete alternatives during the simulated ride experience.
Zielasko et al.\ employed Cybersicker to investigate subjective discomfort, cognitive performance, and physiological stress following extended VR exposure~\cite{Zielasko2026}.
Their results demonstrated increases in both subjective and physiological stress measures, alongside declines in working memory performance following VR exposure.

Due to Cybersicker's open-source availability, verification through in-lab studies, and growing adoption within the VR research community, we decided to leverage it for the work presented in this paper.
In this work, we extended the framework for remote data collection, allowing us to recruit more diverse participant populations, thereby supporting broader validation and generalizability of its findings.

\subsection{In-Person Versus Remote User Studies}
Researchers have traditionally gathered participant data through in-lab user studies conducted at universities and research institutes, as these are often considered the gold standard for controlled data collection~\cite{Roberts2025}. 
Although resource-intensive, requiring expenses such as personnel, facilities, and participant accommodations~\cite{Rupert2017}, in-lab studies offer several advantages, including richer participant-researcher interactions, improved observation of non-verbal behaviors, and access to specialized technologies that may otherwise be unavailable to participants~\cite{Krouwel2019,Hart2023,Souza2024,Dodds2020}.

To complement these approaches, researchers have increasingly explored remote study designs over the years~\cite{Novick2008,Rodriquez2023ci}.
In the late 1990s, psychologists studying human perception~\cite{Loomis1999} and social interaction~\cite{Blascovich_SocialPsychVR2002} proposed the use of VR as a remote research tool, arguing that it would aid scientific replication, enable larger and more representative participant samples, and bridge the gap between experimental control and mundane realism.
As technology matured, computer scientists began to study the process of building and conducting these remote VR studies, for example having participants use smartphones with the Samsung Gear VR~\cite{Steed2016}, recruiting consumers who owned their own VR headsets through platforms such as Mechanical Turk~\cite{Ma2018,Radiah2021}, or running studies remotely to solve logistical issues during the pandemic~\cite{MottelsonSelfAdmin2021}.
Through their adoption, remote studies have been found to cost less, enable recruitment from more diverse populations, and reduce participant anxiety and researcher-participant power differentials~\cite{Buhrmester2018,Lourenco2020,McCoyd2006,Sturges2004,Desai2024}. 
Furthermore, early evidence suggests that study modality has limited influence on observed effect sizes, supporting remote studies as a viable complement for data collection in many research contexts~\cite{Chuey2024}.

When conducting cybersickness studies in person, researchers are often required by Institutional Review Boards (IRBs) to keep participants on-site following study completion to mitigate potential travel-related safety risks, as cybersickness symptoms can persist beyond the study session~\cite{Woo2023-hz}. 
Remote studies offer a promising alternative by enabling participants to complete cybersickness-inducing experiences from their homes, eliminating the need for post-study travel and its associated safety concerns~\cite{Chen2019-texting}. 
However, despite the widespread adoption of remote methodologies in other research domains, we are unaware of prior work that has leveraged remote study designs to investigate cybersickness. 
The recent development of VERA, a platform for conducting remote human-subjects research in VR, provides an opportunity to explore this largely unexamined research space. 
In this work, we leverage VERA to investigate the feasibility of remote cybersickness research.

\section{The VERA Platform}
\label{sec_vera}

We collected all data through VERA, a cloud-based platform for conducting human-subjects VR research with remote, unsupervised participants on their own consumer HMDs\footnote{Platform: \url{https://vera-xr.io}, Project: \url{https://vera-xr.org}}.
VERA comprises two components: a web application for study design and data management, and a Unity plugin for building and running the VR experiment itself.
The \emph{web application} is where researchers define a study---its experimental conditions, the surveys and questionnaires shown to participants, and the distribution method (remote, in-person, or hybrid)---and where they later review and download the collected data.
The \emph{Unity plugin} is used to connect the VR experiment to the platform and instrument data collection into the events and UI within it; at runtime it handles data logging, condition assignment, survey presentation, and session management according to the configuration defined in the web application.
Researchers deploy a study by building and uploading it from the Unity editor, where it is published as a WebXR application and assigned a unique distribution URL.
Participants open their generated URL in the browser on their own headset and complete the study without direct researcher involvement.
The WebXR application is sent directly to the participant's headset from VERA's cloud services, allowing the experiment application to be run directly from the web without requiring any additional download or installation. 
Rendering computations are handled directly on the participant's headset, reducing cloud traffic and mitigating any cloud computation concerns.

\subsection{Positioning}
A recent scoping review of open-source VR research software distinguishes narrow \emph{toolkits} and engine-level \emph{frameworks} from \emph{integrated research platforms} that bundle the experiment lifecycle into a single end-to-end system \cite{Thomsen2026}.
VERA is an integrated research platform for remote, unsupervised, crowd-recruited studies.
Note that VERA is not among the review's entries because the review predates VERA's release and catalogs open-source software only.
Its closest relatives are RemoteLab, a Unity toolkit for supervised remote sessions that the experimenter runs and observes over video conferencing \cite{Lee2022}, and Ouvrai, which supports unsupervised WebXR studies but requires each research group to host its own backend \cite{Cesanek2024}.
All remote tools surveyed in the review are self-hosted in this way~\cite{Thomsen2026}.
VERA instead provides experiment design, WebXR deployment, secure storage, and recruitment integration as a single centrally hosted service, so a lab does not need to assemble or operate its own infrastructure.
Engine-level frameworks such as the Unity Experiment Framework \cite{Brookes2020} structure trials, blocks, and conditions within a single build, a role VERA's plugin covers with its own trial workflow.
Analysis and visualization stages, emphasized by several reviewed platforms, remain future work for VERA.

\subsection{Design}
Two decisions shape the researcher workflow.
First, separating experiment \emph{definition} (the web application) from experiment \emph{implementation} (the Unity plugin) enables the researcher to configure and revise the methodology (including conditions, assignment policy, and instruments) without modifying VR code, and keeps the collected-data schema consistent across studies.
Second, deploying to WebXR does not require per-participant installation. Participants only need to open a URL in their headset browser, without going through an app store, sideloading, or manual file transfer.
The alternative (a native Android build sideloaded onto Quest headsets) would require every participant to register as a developer, which is infeasible for large remote samples.
Additionally, VERA structures the participant session as a sequence that runs without researcher intervention to handle consent, pre- and post-session questionnaires, and the VR experience.
For data collection, VERA writes structured CSV records, each stamped with a participant identifier and a timestamp.
Records are uploaded incrementally throughout a session with retry and backoff, and data is cleared from local storage only after the server confirms receipt.

Collected data is held in a secure, access-controlled cloud repository that operates under its own IRB approval covering the platform's storage and handling of participant data; individual studies, including the present one, additionally carry their own approval (Section~\ref{sec_method}).
For recruitment through Prolific, the distribution URL includes the participant's Prolific identifier as a URL parameter, and on completion VERA returns the study's Prolific completion code so the session can be marked for payment.
Records thus do not have directly identifying information---each participant is keyed only by that pseudonymous Prolific identifier, under which the research team accesses the data.

\section{Method}
\label{sec_method}

In this section, we describe the methodology we used for the evaluation in this paper.
The study was approved by the IRB of the University of Central Florida (\# STUDY00008810).
Informed consent was obtained from all participants.

\subsection{Study Design}

We conducted an observational, remote study of cybersickness deployed fully online through the VERA platform (Section~\ref{sec_vera}) and the Prolific participant recruitment service.
Participants completed the study unsupervised, in a location of their choosing, on their own consumer HMD.
We adapted the protocol from an established in-lab cybersickness testbed study \cite{zielasko2024cybersicker,Zielasko2026} so that the core stimulus and measures align with the in-lab setup, which supports a comparison between the two studies (Section~\ref{sec_results}).
The remote deployment enables a substantially larger and more demographically diverse sample than is feasible in a controlled laboratory.
The study therefore had two complementary statistical objectives.
First, we evaluated whether the direction, temporal pattern, and magnitude of key cybersickness outcomes observed through VERA were broadly consistent with those reported in the reference in-lab study.
Second, we evaluated whether the larger and more heterogeneous VERA cohort provided sufficient precision to characterize associations between cybersickness and participant-level factors that could not be estimated reliably in the smaller laboratory cohort.
These two objectives motivate a three-cohort design.
We compare the full remote sample against the in-lab reference sample, whose data the original authors shared with us, and against a demographically matched subset of the remote sample built to mirror the in-lab sample.
Section~\ref{sec:mirror_cohort_selection} defines these cohorts and describes the matching procedure.

\paragraph*{\textbf{Comparability}}
We designed our study based on the reference in-lab study by Zielasko et al.~\cite{Zielasko2026}, deviating from its protocol in two respects.
First, the original study collected a variety of physiological measures (e.g., saliva samples) that were central to its research questions but were not relevant to our study or impractical to collect remotely.
Omitting them let us administer the post-exposure SSQ directly after the experience instead of roughly 15 minutes later.
Second, we set the termination criterion of the 0--10 FMS to the maximum of 10 instead of the 7 used in the original study, as we wanted to extend the research into the higher severity regions that are often prevented by procedural constraints of in-lab participation.
We discuss potential effects of these differences in Section~\ref{sec_discussion_divergence}.

\paragraph*{\textbf{Remote Data Collection with VERA}}
We deployed the study through the VERA infrastructure described in Section~\ref{sec_vera} and used it to record every participant session.
For this study, the logs captured the FMS and SSQ responses, as well as session timing and interaction events, which were stored as de-identified records keyed by a study participant identifier.
Recruitment, completion tracking, and payment were handled through Prolific.

\vspace{20pt}
\subsection{Stimuli \& Material\label{stimulus}}

\begin{figure*}
\centering
\begin{subfigure}[b]{0.329\textwidth}
    \includegraphics[width=\linewidth]{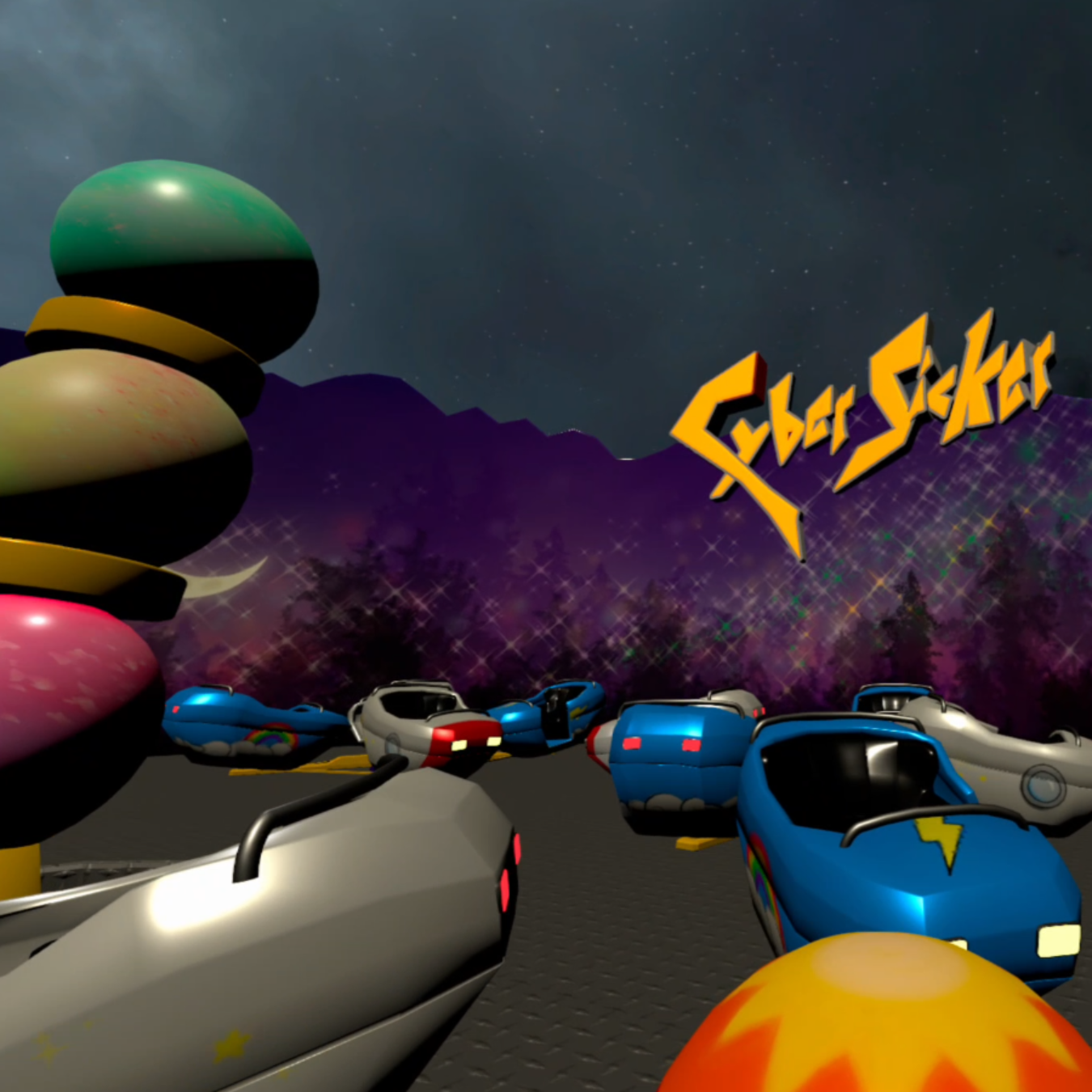}
    \caption{}
\end{subfigure}
\hfill
\begin{subfigure}[b]{0.329\textwidth}
    \includegraphics[width=\linewidth]{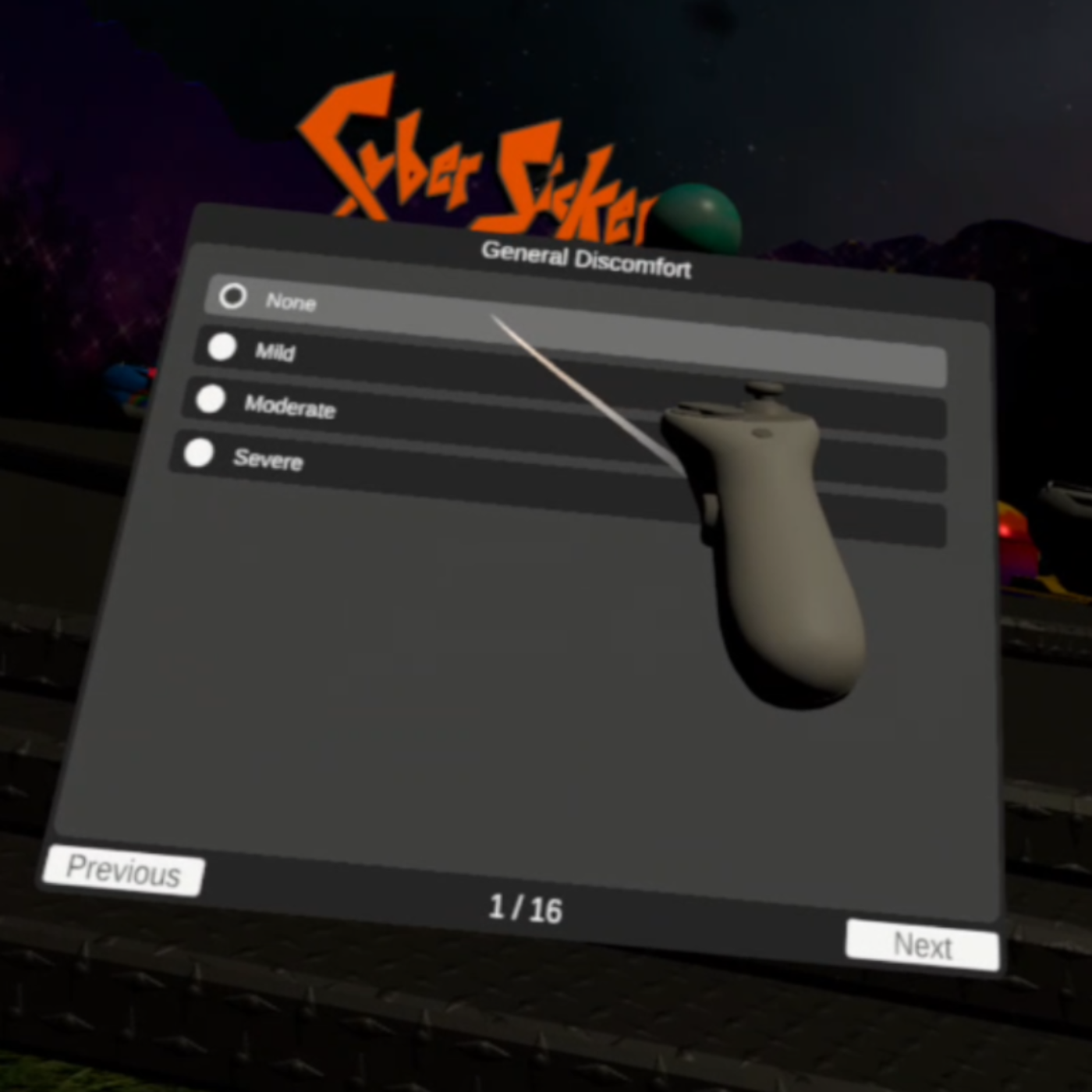}
    \caption{}
\end{subfigure}
\hfill
\begin{subfigure}[b]{0.329\textwidth}
    \includegraphics[width=\linewidth]{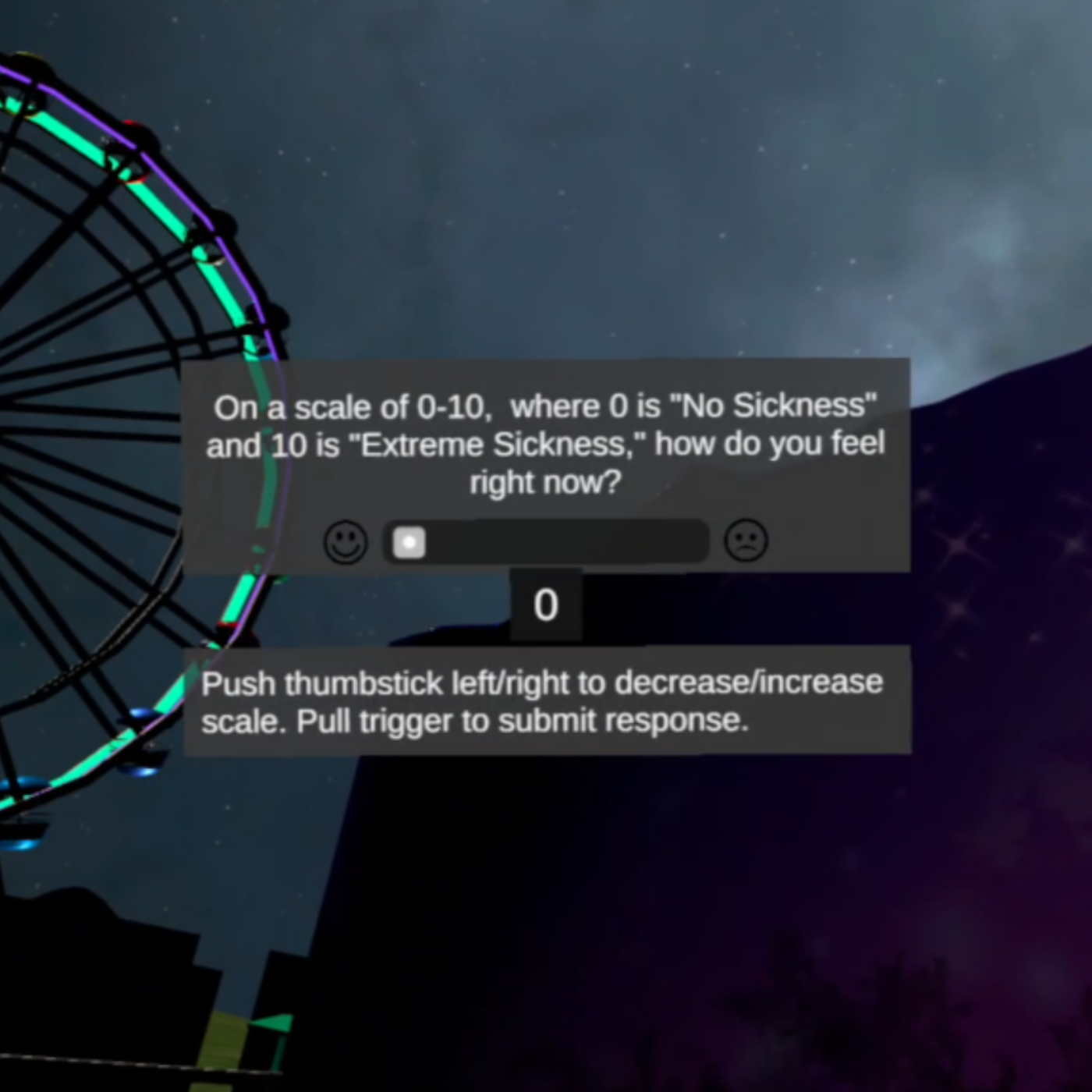}
    \caption{}
\end{subfigure}
\caption{Visual stimuli used in the experiment: (a) seated in a cart of the Cybersicker amusement park ride, (b) completing the pre- and post-exposure SSQ, and (c) completing the FMS in regular intervals during the ride.}
\label{fig_cybersicker}
\end{figure*}

For inducing cybersickness, we used the open-access simulator \emph{Cybersicker}, which was developed by Zielasko and Law~\cite{zielasko2024cybersicker} (cf.\ Section~\ref{sec_rw_testbeds}).
Participants in this simulator experience a seated amusement park ride that was designed to provide a stimulus similar to a virtual roller coaster ride~\cite{Ang2023}, but provides more experimental control while maintaining a realistic experience.
This simulator is implemented in Unity and based on a real amusement park ride known as the \textit{Breakdance}\footnote{\url{https://en.wikipedia.org/wiki/Breakdance_(ride)}}.
Participants in this study were seated in a cart of the simulated ride with no virtual body representation.
In this simulator, each cart is grouped with three others around a central pivot point, while individual carts can rotate around their yaw and pitch axes.
All cart groups on the tilted platform collectively rotate around the platform's central axis around a central tower structure.
Figure~\ref{fig_cybersicker} shows the virtual experience from a participant's point of view.

During the simulation, both the participant's own cart and gondolas in the surrounding virtual fairground began to rotate.
They gradually increased in speed automatically every 30 seconds, following the protocol of the in-lab study~\cite{Zielasko2026}.
Participants' task was to experience and observe the ride passively, while they could look around freely in the virtual scene.

Because the experience is rendered locally on each participant's own headset (Section~\ref{sec_vera}), its performance depends on the device.
We therefore tested the experience on all three supported headsets, the Meta Quest 2, Quest 3, and Quest 3S, before deployment, and confirmed that it ran at a consistent frame rate and presented an equivalent visual experience across the devices.

\subsection{Procedure}

We recruited participants through the Prolific platform.
Prolific pre-screening filters restricted recruitment to adults who reported owning a compatible VR headset, and the recruitment notice showed the study title, description, estimated duration, and pay rate.
Eligibility was limited to adults between 18 and 65 years of age who owned a Meta Quest 2, Quest 3, or Quest 3S headset, had normal or corrected-to-normal vision, and were able to sit comfortably for 30 minutes.
We excluded participants with a history of epilepsy or seizures, vestibular or motor disorders, mild cognitive impairment or dementia, and participants who were pregnant.

Participants who accepted the study were redirected to a Qualtrics questionnaire whose first page embedded a downloadable consent form.
Participants reviewed the consent form and either gave explicit consent by selecting ``I consent and wish to participate'' or declined and were returned to Prolific.
Participants were informed in the Prolific recruitment notice and in the consent form that the experiment aimed to induce cybersickness.

After consenting, participants first completed the Qualtrics pre-questionnaire (approximately 10 minutes).
The questionnaire included attention checks to screen for non-attentive responding.
They then donned their HMD and opened the VR experiment through a WebXR link to complete the remainder of the study in VR.
In the VR portion of the study, participants completed the SSQ for the first time, assessing their current symptoms related to cybersickness.
The experiment then introduced the FMS scale via slides, explaining that a score of 0 represented no symptoms, while a score of 10 indicated severe sickness, such as nausea leading to vomiting.
Participants were informed that during the sickness stimulus, they would be asked every 30 seconds to report their current sickness level via FMS.
If they reported a score of 10, the experiment would be stopped immediately, and they were made aware of this.

The VR stimulation lasted either for a maximum of eight minutes or until the participant reached the termination threshold.
Participants then completed the post-exposure SSQ in VR.
After completing the post-exposure SSQ, participants removed their headset and received compensation through Prolific.

\subsection{Measures\label{measures}}

In this section, we describe the measures we included in the experiment.

\paragraph*{\textbf{Demographic Questionnaire}}
As part of the pre-questionnaire participants completed via Qualtrics, we collected core demographic information, including age, sex, height and weight (from which we computed body mass index), prior VR experience and familiarity, subjective susceptibility to motion sickness, and the participant's expectation of becoming sick during the experience.

\paragraph*{\textbf{Visually Induced Motion Sickness Susceptibility Questionnaire (VIMSSQ)}}
Participants completed the nausea subscale of the VIMSSQ (VIMSSQ-N) via Qualtrics~\cite{Keshavarz2023vimssq}.
This instrument evaluates the susceptibility of participants to \acf{VIMS}, that is, symptoms related to motion sickness while using visual displays.
Participants rated how often they had experienced nausea when using each of the instrument's 11 visual device and display types (Never, Rarely, Sometimes, or Often, with an N/A option for devices they have not used).
Item responses are coded 0 to 3, summed over answered items, and normalized to a 0 to 33 subscale score.

\paragraph*{\textbf{Simulator Sickness Questionnaire (SSQ)}}
Participants filled out the SSQ~\cite{Kennedy1993} in VR using their hand-held controller (see Figure~\ref{fig_cybersicker}b), both before and after the VR stimulation.
This instrument measures symptoms of cybersickness on a 4-point scale from None to Severe using 16 items.
It provides the three subscales of nausea, disorientation, and oculomotor as well as an overall score of cybersickness.

\paragraph*{\textbf{Fast Motion Sickness Scale (FMS)}\label{sec_fms}}
Participants were asked to rate their feelings of comfort throughout the 8-minute VR stimulation on a scale from 0 (not sick at all) to 10 (extreme sickness).
This scale is an adapted version of the FMS~\cite{Keshavarz2011validating}.
We adapted the original 0 to 20 range to an 11-point 0 to 10 scale to make it more intuitive for the participants under the potentially stressful circumstances this experiment evokes~\cite{zielasko2016hmdnav, Adhikari2022}, matching the scale used in the in-lab testbed protocol \cite{zielasko2024cybersicker}.
The FMS was used during the experiment to evaluate the current well-being of the participants during the VR stimulation.
For this purpose, participants reported their current sickness every 30 seconds during the motion stimulation using their hand-held controller (see Figure~\ref{fig_cybersicker}c).
Once they reached the threshold of 10, the stimulation was terminated to protect the participants from severe physical reactions such as throwing up.

\subsection{Participants\label{participants}}

We recruited participants remotely through Prolific\footnote{\url{https://www.prolific.com}} under the eligibility criteria described above.
Of those who agreed to participate, 912 completed the pre-experiment questionnaire, which collected demographics and the susceptibility and expectation measures.
A total of 381 participants launched the VR experience through VERA, of whom 288 completed the full session, providing FMS ratings and both the pre- and post-exposure SSQ.
Of these, 277 could be linked to a completed questionnaire.
The remaining 11 could not be matched to their questionnaire responses because of a missing or malformed participant identifier, so their demographic data were unavailable.
From the 277, we removed 10 participants who failed a data-quality check, either an explicit attention-check question or a demographic response that was out of range or impossible (including one age value in the tens of thousands, and an impossible body mass index), leaving 267.
We additionally excluded four participants whose response on a modeled variable could not be used, two who declined to report their sex and two who marked every device category of the VIMSSQ-N as not used (which is unscoreable based on the questionnaire's rules~\cite{Keshavarz2023vimssq}), for a final analysis sample of 263.
Table~\ref{tab:recruitment_funnel} summarizes this recruitment funnel.
Overall, 263 of the 912 participants who completed the pre-experiment questionnaire entered the final analysis sample (28.8\%).
Each participant gave informed consent and was compensated at a rate of \$15 per hour, pro-rated for the time spent in the study (approximately \$7.50 for the 30-minute session), paid through Prolific.

\begin{table}[t]
\centering
\caption{Recruitment funnel for the remote VERA deployment. Each percentage is relative to the previous stage. 
}
\label{tab:recruitment_funnel}
\begin{tabular}{lrr}
\toprule
Stage & N & \% of previous \\
\midrule
Completed pre-questionnaire & 912 & --- \\
Launched the VR experience & 381 & 41.8\% \\
Completed full VR session & 288 & 75.6\% \\
Linked to a questionnaire & 277 & 96.2\% \\
Passed quality screening & 267 & 96.4\% \\
Final analysis sample & 263 & 98.5\% \\
\bottomrule
\end{tabular}
\vspace{-8pt}
\end{table}

Of the 93 participants who launched the experience but did not complete the full session, 81 exited before the motion stimulation began.
The remaining 12 entered the ride and provided FMS ratings but did not submit the post-exposure SSQ.
For 8 of the 12, the ride ended by protocol, either after the full duration or at the FMS ceiling, so only 4 participants quit during the motion stimulation itself.

\subsubsection{Participant Cohorts}
\label{sec:mirror_cohort_selection}
We compared our results against a related in-lab study by Zielasko et al.~\cite{Zielasko2026} and analyzed demographic factors related to cybersickness in a large population.
Three cohorts support these analyses: the reference sample from the in-lab study, the full remote sample, and an individually matched subset of the remote sample.
We refer to these cohorts as the \acf{IPC}, \acf{VFC}, and \acf{VMC}, respectively.
A direct comparison on the \ac{VFC} conflates differences in cohort size, demographic composition, and protocol.
To account for these differences, we constructed the \ac{VMC} by matching each \ac{IPC} participant to one VERA participant with a similar demographic profile.

\paragraph*{\textbf{VERA Full Cohort (VFC)}}
The \ac{VFC} comprised 263 participants who passed the inclusion requirements and attention checks described in Section~\ref{participants}. The sample was balanced by sex (130 male, 133 female). Ages ranged from 18 to 65 years, with a mean of 34.2 years (SD 10.7) and a median of 33. Mean BMI was 29.5 kg/m² (SD 8.0), at the boundary between overweight and obese. Because remote recruitment required ownership of a Meta Quest headset, participants were predominantly experienced VR users: median self-rated VR familiarity was 6 on a 1 to 7 scale (IQR 5 to 7), with 169 of 263 in the highest familiarity band and only two in the lowest. Nausea susceptibility (VIMSSQ-N) was $M = 5.86$ ($SD = 5.72$) on the 0 to 33 scale. Sickness expectation was mixed, with most participants not expecting to feel sick (139 ``no,'' 78 ``yes,'' 46 ``don't know'').

\paragraph*{\textbf{In Person Cohort (IPC)}}
The \ac{IPC} comprised 30 in-lab participants, balanced by sex (15 male, 15 female). It was a narrow band of young adults: mean age 24.9 years (SD 4.6, median 24.5), ranging from 19 to 44 years old and concentrated in the mid-twenties, consistent with a university convenience sample. Mean BMI was 24.8 kg/m² (SD 4.0), in the normal-weight range. In contrast to the \ac{VFC}, \ac{IPC} participants were almost entirely new to VR: median VR usage/familiarity was 1 on the study's 1 to 5 scale (IQR 1 to 2), with 27 of 30 in the lowest familiarity band and none in the highest. Nausea susceptibility (VIMSSQ-N) was $M$\,=\,3.93 ($SD$\,=\,4.66), and sickness expectation was distributed across all three levels (11 ``no,'' 7 ``yes,'' 12 ``don't know'').

\paragraph*{\textbf{VERA Mirror Cohort (VMC)}}
We matched the \ac{IPC} and \ac{VMC} data on six variables shared between studies: sex, age, BMI, VR familiarity, nausea susceptibility (VIMSSQ-N), and sickness expectation.
Sex, age, nausea susceptibility, and BMI are directly comparable across studies.
The remaining two constructs, VR familiarity and sickness expectation, use different response scales between studies and require explicit binning to a shared three-level scale.
\suppalt{Section~\ref{sec:supp_construct_mapping} in the supplemental document lists the source columns and the bin definitions.}{The supplemental material lists the source columns and the bin definitions.}

To form the \ac{VMC}, for each \ac{IPC} participant, our matching algorithm searched the VERA pool for candidates that shared the \ac{IPC} profile on every active constraint and sampled one candidate uniformly at random.
Sampling was without replacement across the 30 \ac{IPC} participants, so the \ac{VMC} contained 30 distinct VERA participants.
When no candidate matched the strictest constraint set, the algorithm relaxed constraints in an ordered series of tiers and retries.
\suppalt{Section~\ref{sec:supp_matching_algorithm} in the supplemental material provides the full tier ordering and the rationale for the relaxation order.}{The supplemental material provides the full tier ordering and the rationale for the relaxation order.}

The composition of the \ac{VMC} depends on the random number generator seed used in the matching candidate selection.
We ran the matcher for 1000 candidate seeds and used the seed whose cohort showed the best covariate balance with the \ac{IPC} as the canonical \ac{VMC} for all comparisons reported here.
No outcome measures (FMS or SSQ) informed the selection.
To quantify the remaining seed dependence, we recomputed every \ac{VMC}-versus-\ac{IPC} comparison for all 1000 candidate cohorts.
\suppalt{In the supplemental material, Section~\ref{sec:supp_seed_distribution} details the balance score and reports these distributions, and Section~\ref{sec:supp_match_quality} reports the per-pair match quality at the chosen seed.}{The supplemental material details the balance score, reports these distributions, and reports the per-pair match quality at the chosen seed.}

The resulting \ac{VMC} matched the \ac{IPC} exactly on sex (15 male, 15 female) and closely on the continuous variables, differing on average by 1.7 years in age, 2.2 BMI points, and 1.1 VIMSSQ-N points per pair.
The sickness-expectation bin agreed in 27 of 30 pairs.
VR familiarity was the one construct that could not be matched (bin agreement in none of the 30 pairs), because the remote pool of headset owners contained almost no VR-naive candidates.
Overall, the \ac{VMC} mirrored the \ac{IPC} on all shared axes except prior VR exposure.

\paragraph*{\textbf{Sample-size considerations}}
The precision of the cross-study comparison is constrained primarily by the size of the \ac{IPC} sample (N=30).
The \ac{VFC} includes 263 participants and five prespecified participant-level predictors, approximately 50 observations per predictor.
Under a conventional linear-model approximation, this sample provides approximately 80\% power to detect a small incremental predictor effect of Cohen's $f^2 \approx 0.03$ at a two-sided significance level of 0.05.
The ordered-beta and mixed-effects models have more complex variance structures, so we treat this calculation as an approximate sensitivity benchmark.
The principal statistical advantage of the larger cohort is increased precision and demographic coverage, which allows estimation of participant-level associations that cannot be reliably characterized by a 30-participant sample.
Recruitment continues beyond the present analysis, with a planned final sample exceeding 1{,}000 participants.
The expanded sample is intended to support higher-precision estimation, nonlinear and interaction analyses, hardware-specific comparisons, and independent validation of predictive models.

\vspace{12pt}
\section{Results}
\label{sec_results}

\begin{table*}[ht]
\centering
\caption{Demographic, susceptibility, FMS, and SSQ measures for the \ac{IPC}, the demographically matched \ac{VMC} (Section~\ref{sec:mirror_cohort_selection}), and the \ac{VFC}. Cells show $M \pm SD$ unless noted, median [IQR] for ordinal or right-skewed measures, and counts (\%) for proportions. The FMS max-out rows use each study's protocol ceiling (FMS $\geq$ 7 in the \ac{IPC}, FMS $=$ 10 in VERA).}
\label{tab:IP_vera_comparison}
\begin{tabular}{llll}
\toprule
Measure & \ac{IPC} (N=30) & \ac{VMC} (N=30) & \ac{VFC} (N=263) \\
\midrule
\multicolumn{4}{l}{\textbf{Demographics \& Susceptibility}} \\
Age (years) & 24.93 $\pm$ 4.58 & 25.93 $\pm$ 4.65 & 34.22 $\pm$ 10.72 \\
Sex (M / F) & 15 / 15 & 15 / 15 & 130 / 133 \\
BMI (kg/m$^2$) & 24.82 $\pm$ 4.01 & 23.85 $\pm$ 4.01 & 29.50 $\pm$ 8.04 \\
VR usage / familiarity & 1.0 [1.0, 2.0] & 6.0 [6.0, 7.0] & 6.0 [5.0, 7.0] \\
VIMSSQ-N nausea susceptibility (0--33) & 3.93 $\pm$ 4.66 & 3.86 $\pm$ 4.46 & 5.86 $\pm$ 5.72 \\
\midrule
\multicolumn{4}{l}{\textbf{FMS}} \\
Max FMS (peak) & 6.00 $\pm$ 1.36 & 6.50 $\pm$ 3.36 & 7.14 $\pm$ 3.64 \\
Mean FMS & 3.18 $\pm$ 0.96 & 3.58 $\pm$ 2.18 & 3.82 $\pm$ 2.30 \\
FMS AUC (FMS$\cdot$s) & 1069 $\pm$ 441 & 1187 $\pm$ 750 & 1101 $\pm$ 784 \\
Final FMS (last reading) & 5.97 $\pm$ 1.38 & 6.23 $\pm$ 3.59 & 6.90 $\pm$ 3.83 \\
Time to FMS max out (s, hit only) & 275 [230, 320] & 270 [150, 390] & 240 [180, 330] \\
Hit FMS max out & 16 / 30 (53.3\%) & 10 / 30 (33.3\%) & 130 / 263 (49.4\%) \\
\midrule
\multicolumn{4}{l}{\textbf{SSQ (pre $\to$ post)}} \\
$\Delta$SSQ Total & 33.5 $\pm$ 29.2 & 54.0 $\pm$ 41.4 & 57.6 $\pm$ 41.2 \\
$\Delta$SSQ Nausea & 37.5 $\pm$ 32.0 & 55.3 $\pm$ 41.2 & 56.0 $\pm$ 40.5 \\
$\Delta$SSQ Oculomotor & 20.0 $\pm$ 20.0 & 31.1 $\pm$ 26.7 & 32.9 $\pm$ 26.8 \\
$\Delta$SSQ Disorientation & 33.4 $\pm$ 34.2 & 63.1 $\pm$ 53.2 & 72.2 $\pm$ 57.7 \\
\bottomrule
\end{tabular}
  \vspace{-10pt}
\end{table*}

In this section, we first present results from our comparison of the VERA participant data to the results obtained in the \ac{IPC}, addressing the first statistical objective.
We then present further findings from the larger and more diverse \ac{VFC}, addressing the second objective.

\subsection{Cybersickness Outcomes: VERA Versus In-Lab Study}
\label{sec:cybersickness_outcomes}

Here we compare cybersickness outcomes across three cohorts: the \ac{IPC} (N=30), the demographically matched \ac{VMC} (N=30), and the \ac{VFC} (N=263).
Summary statistics for every measure appear in Table~\ref{tab:IP_vera_comparison}.

\begin{figure}[t]
  \centering
  \includegraphics[width=\linewidth]{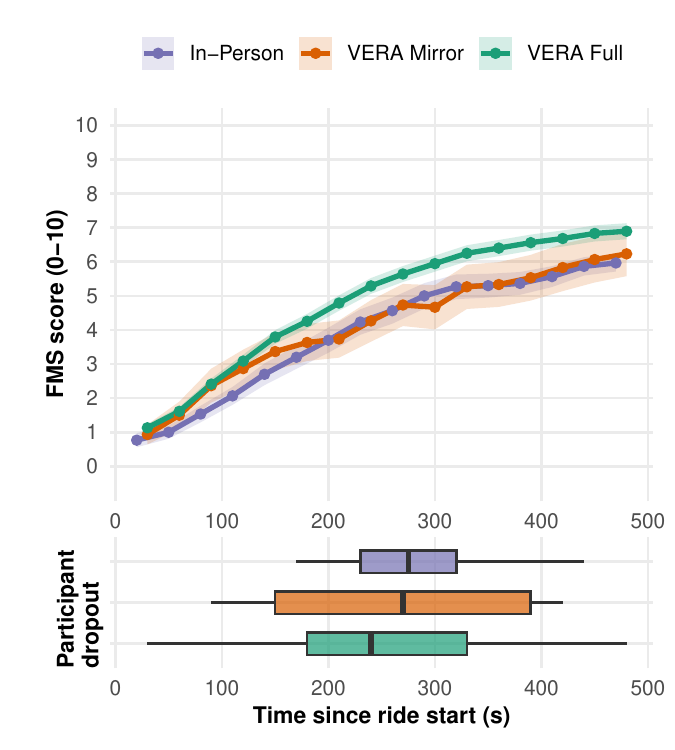}
  \vspace{-13pt}
  \caption{Mean FMS over the ride by cohort (mean $\pm$ SE; top), with the distribution of time to reach the FMS ceiling for participants who reached it (bottom), on a shared time axis.
  Boxes show the median and interquartile range.
  16 of 30 \ac{IPC}, 10 of 30 \ac{VMC}, and 130 of 263 \ac{VFC} participants reached the ceiling.
  Ceilings follow each study's protocol (\ac{IPC}: FMS $\geq$ 7, VERA: FMS $=$ 10).
  Participants who reached their ceiling and stopped early have their last FMS value carried forward to the end of the ride.}
  \label{fig:fms_traj_maxout_marginal}
  \vspace{-5pt}
\end{figure}

\begin{figure}[t]
  \centering
  \includegraphics[width=0.8\linewidth]{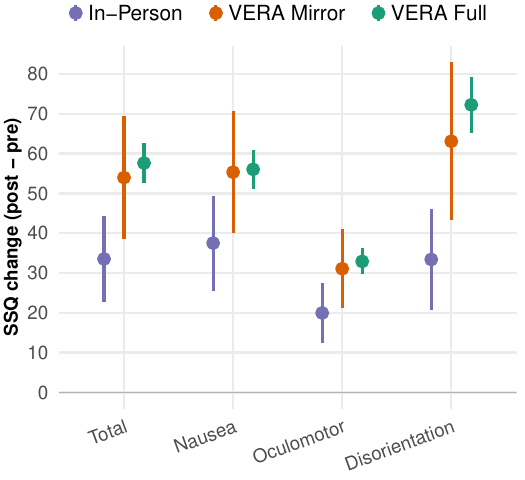}
  \caption{SSQ change after VR exposure (post minus pre), by cohort: mean and 95\% confidence interval for the total score and the three subscales.}
  \label{fig:ssq}
\end{figure}

Symptom onset was rapid in every cohort.
Among participants who reached their ceiling, the median time to reach it was 275\,s in the \ac{IPC}, 270\,s in the \ac{VMC}, and 240\,s in the \ac{VFC}, with overlapping interquartile ranges (Figure~\ref{fig:fms_traj_maxout_marginal}, bottom).
Between 33\% and 53\% of each cohort reached the ceiling (16 of 30 in the \ac{IPC}, 10 of 30 in the \ac{VMC}, and 130 of 263 in the \ac{VFC}).
Mean FMS rose steadily over the eight-minute ride in all three cohorts (Figure~\ref{fig:fms_traj_maxout_marginal}, top).
With early terminations held at their last value, the final group mean reached 5.97 in the \ac{IPC}, 6.23 in the \ac{VMC}, and 6.90 in the \ac{VFC}.
The remote VERA cohorts tracked the \ac{IPC} trajectory closely throughout the ride.

SSQ confirmed substantial sickness in every cohort.
From a near-zero baseline, SSQ rose sharply after exposure ($\Delta$SSQ Total 33.5 in the \ac{IPC}, 54.0 in the \ac{VMC}, and 57.6 in the \ac{VFC}; Table~\ref{tab:IP_vera_comparison}, Figure~\ref{fig:ssq}).
Disorientation was the largest subscale in the VERA cohorts, whereas nausea was the largest in the \ac{IPC}, and oculomotor symptoms were the smallest contributor in every cohort (Figure~\ref{fig:ssq}).
The VERA cohorts reported somewhat higher SSQ than the \ac{IPC}, which is consistent with VERA's higher FMS ceiling (10 versus 7), which allows longer exposure before the ride stops.

Taken together, the demographically matched \ac{VMC} produced a cybersickness profile broadly comparable to the \ac{IPC} across onset timing, FMS progression, and SSQ severity.
These comparisons do not depend on the particular matched cohort; \suppalt{Table~\ref{tab:seed_distribution} in the supplemental material}{the supplemental material} reports their distribution across 1000 candidate \ac{VMC} draws.
The \ac{VFC} extends these observations to a larger and more demographically varied sample.

\subsection{Modeling Participant Factors}
\label{sec:factor_modeling}

We next conducted an exploratory analysis to examine how participant factors relate to cybersickness.
We modeled five factors available in both studies: age, sex, VR familiarity, nausea susceptibility (VIMSSQ-N), and sickness expectation.
We fit one model per cohort, with all five factors entered together, so each factor's estimate is adjusted for the others.
We focus on two primary outcomes, one per symptom family: mean FMS over the ride and the pre-to-post SSQ change.
We also fit the FMS time course as a mixed model.
Three further outcomes (time to reach the ceiling, whether the participant reached it, and post-exposure SSQ total) are reported descriptively in Section~\ref{sec:cybersickness_outcomes} and Table~\ref{tab:IP_vera_comparison}.

We matched the distributional family of each model to the distribution of its outcome.
Mean FMS is bounded on the 0--10 scale, so we modeled it as an ordered-beta response.
Coefficients from this model are on the logit scale, so we interpret their sign and significance directly. Effect figures back-transform predictions to the 0--10 FMS scale to show magnitude.
The SSQ change score can be negative, so we used a Gaussian model.
We fit the FMS time course as a linear mixed model with a per-participant random intercept and slope, the only model here with random effects; the participant-level outcomes have one observation per person and therefore no within-cohort grouping.

We fit all models by maximum likelihood and report coefficients with 95\% confidence intervals and Wald tests.
Each factor enters with a single degree of freedom, so each coefficient's test is the test for that factor.
Residual diagnostics (simulation-based residuals, dispersion, and zero-inflation checks) showed no evidence of misfit, and collinearity among the five factors was low (all variance-inflation factors below 2).
The \ac{IPC} and \ac{VMC} each have 30 participants and five predictors, so their estimates are exploratory and carry wide intervals.
The \ac{VFC} (N=263) is the well-powered model and is the basis for the inferences below.
\suppalt{Table~\ref{tab:factor_models} in the appendix summarizes the model fit to each measure, and Table~\ref{tab:factor_coefficients} reports the full coefficients (estimates, standard errors, and 95\% confidence intervals) for the two primary outcomes.}{The supplemental material summarizes the model fit to each measure and reports the full coefficients (estimates, standard errors, and 95\% confidence intervals) for the two primary outcomes.}

\begin{figure*}[t]
    \centering
    \captionsetup[subfigure]{skip=-5pt}
    \captionsetup{skip=4pt}
     \begin{subfigure}[b]{0.32\textwidth}
       \includegraphics[width=\linewidth]{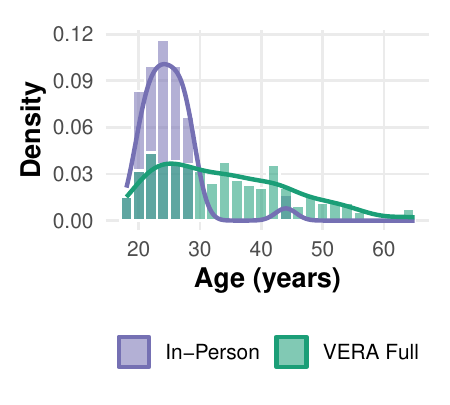}
       \caption{}
       \label{fig:factor_age_dist}
    \end{subfigure}%
    \hfill
  \begin{subfigure}[b]{0.65\textwidth}
    \includegraphics[width=\linewidth]{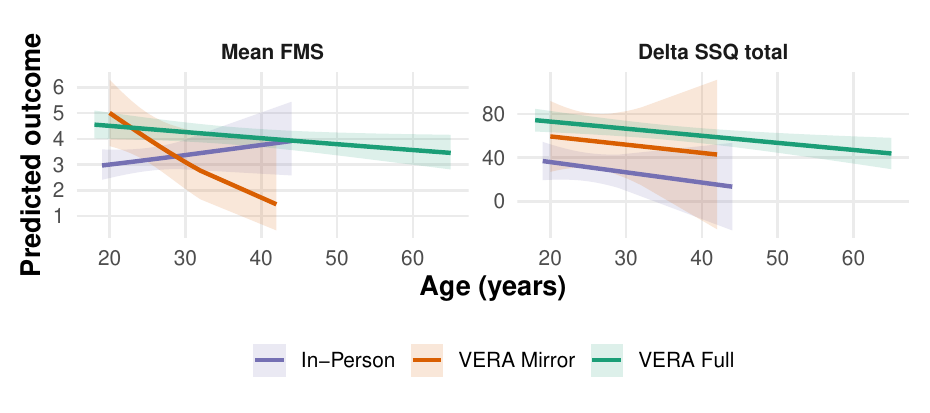}
    \caption{}
    \label{fig:factor_age_effects}
  \end{subfigure}
  \caption{Age by cohort.
  (a) Age distribution of the \ac{IPC} and the \ac{VFC}.
  Bars show within-cohort density and curves show kernel density estimates.
  (b) Model-based effect of age on the two primary outcomes (mean FMS and SSQ change), by cohort.
  Each panel shows the predicted outcome across the observed age range, holding sex, VR familiarity, susceptibility, and expectation at their reference or typical values, with 95\% confidence ribbons.
  Curves are drawn over each cohort's own age range, so the \ac{IPC} curves cover only young adults.
  }
  \label{fig:factor_age}
  \vspace{-22pt}
\end{figure*}

\subsection{Age}
\label{sec:factor_age}

\subsubsection{Age Distributions Differed Between Cohorts}

The \ac{IPC} and the \ac{VFC} differed sharply in age (Figure~\ref{fig:factor_age_dist}).
VERA participants were older: the median age was 33 years in the \ac{VFC} versus 24.5 in the \ac{IPC}, a shift of 8.5 years (Mann--Whitney $p < 10^{-5}$, Cliff's $\delta = -0.54$, a large effect).
VERA participants were also far more spread out: the standard deviation was 10.7 years versus 4.6 in the \ac{IPC}, a variance ratio of 5.5 and an interquartile-range ratio of 3.4 (Levene, Fligner--Killeen, and $F$ tests all $p < 10^{-5}$).
A Kolmogorov--Smirnov test confirmed the overall distributions differ ($p < 10^{-8}$).
The \ac{IPC} was a narrow band of young adults (ages 19--44, concentrated in the mid-twenties), whereas the \ac{VFC} spans the adult range (18--65).

\subsubsection{Age and Cybersickness Outcomes}

The wider age range in the VERA sample exposes an age relationship not shown by the \ac{IPC} (Figure~\ref{fig:factor_age_effects}).
In the \ac{VFC}, older participants reported less sickness after controlling for the other four factors.
Each additional year of age was associated with a lower SSQ change score ($-0.65$ SSQ points per year, 95\% CI $[-1.09, -0.21]$, $p = 0.004$).
Mean FMS showed the same negative association ($b = -0.010$, 95\% CI $[-0.019, -0.001]$, $p = 0.04$).
The \ac{VMC} showed the same negative direction on both outcomes, reaching significance for mean FMS ($p = 0.03$) but not for the SSQ change ($p = 0.69$).

The \ac{IPC} showed no reliable age association on any outcome.
Its age coefficients were small, mixed in sign, and never significant, which is the expected result when age varies over only a narrow range.

\subsection{Sex}
\label{sec:factor_sex}

The cohorts did not differ in sex composition.
The \ac{IPC} was balanced by design (15 male, 15 female), and the \ac{VFC} was 130 male and 133 female, about 50\% male (chi-square $p \approx 1$).

\begin{figure}[t]
  \centering
  \includegraphics[width=\linewidth]{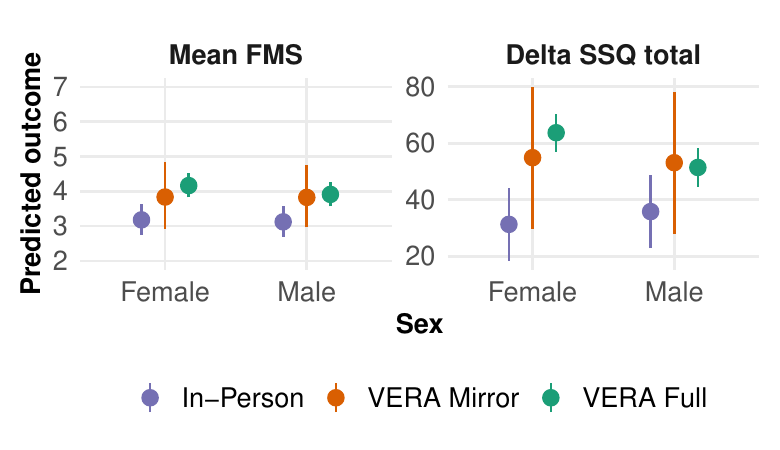}
  \caption{Model-based effect of sex on mean FMS and SSQ change by cohort, with 95\% intervals.
  Predictions hold the other four factors at their reference or typical values.}
  \label{fig:factor_sex_effects}
\end{figure}

In the \ac{VFC}, male participants reported less SSQ than female participants after adjusting for the other factors (Figure~\ref{fig:factor_sex_effects}).
The SSQ change score was 12.3 points lower for male participants (95\% CI $[-22.2, -2.4]$, $p = 0.016$).
Mean FMS did not differ by sex ($p = 0.32$).
The \ac{VMC} showed the same direction for SSQ change without significance, and the \ac{IPC} showed no sex differences.
Female participants therefore tended to report more SSQ than male participants, and this difference was detected only in the well-powered cohort.

\subsection{VR Familiarity}
\label{sec:factor_vrfam}

VR familiarity is the factor on which the two studies overlap least.
\ac{IPC} participants were almost all VR-naive, whereas the \ac{VFC} was predominantly experienced, a large difference (Table~\ref{tab:IP_vera_comparison}; chi-square $p < 0.001$).
The two studies also measured the construct with different instruments (Section~\ref{sec:mirror_cohort_selection}), so the bin mapping is approximate.

\begin{figure}[t]
  \centering
  \includegraphics[width=\linewidth]{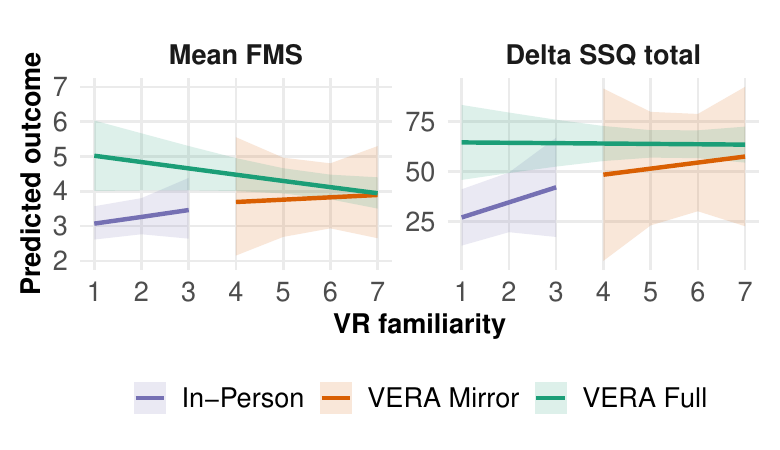}
  \caption{Model-based effect of VR familiarity on the two primary outcomes (mean FMS and SSQ change), by cohort, with 95\% intervals.
  Curves use each cohort's native familiarity scale and own observed range.}
  \label{fig:factor_vrfam_effects}
\end{figure}

Because the \ac{IPC} has almost no variation in VR familiarity, it cannot estimate a VR effect (Figure~\ref{fig:factor_vrfam_effects}).
In the \ac{VFC}, more VR-familiar participants showed a weak trend toward lower mean FMS ($b = -0.07$, $p = 0.09$), with no effect on SSQ change ($p = 0.93$).
Neither primary outcome showed a VR effect in the \ac{VMC}.
VR familiarity was therefore at most weakly protective within VERA and could not be compared against the \ac{IPC}.

\begin{figure*}[t]
  \centering
  \captionsetup[subfigure]{skip=-5pt}
  \captionsetup{skip=4pt}
  \begin{subfigure}[b]{0.49\textwidth}
      \includegraphics[width=\linewidth]{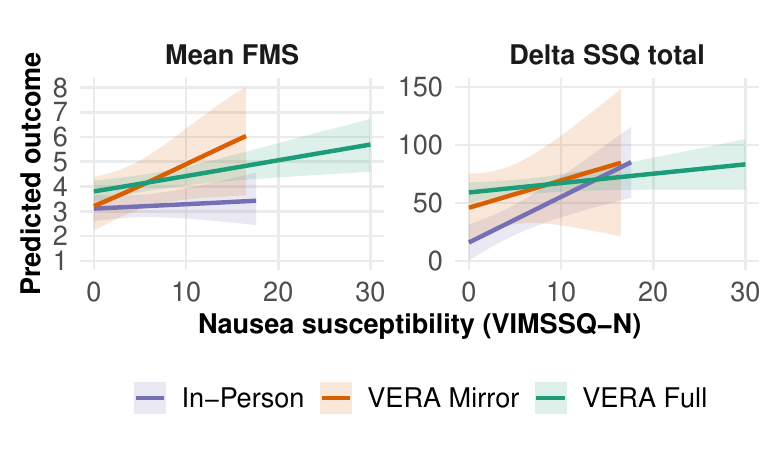}
      \caption{}
      \label{fig:factor_susc_effects}
  \end{subfigure}
  \hfill
  \begin{subfigure}[b]{0.49\textwidth}
       \includegraphics[width=\linewidth]{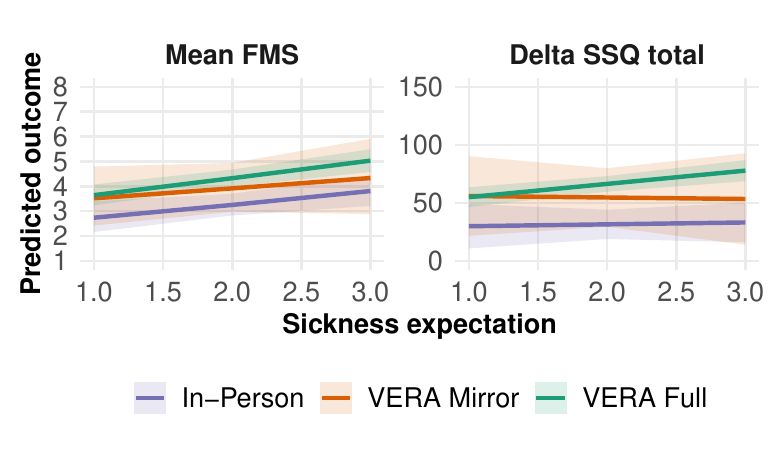}
       \caption{}
       \label{fig:factor_expect_effects}
  \end{subfigure}
  \caption{Model-based effects of (a) nausea susceptibility (VIMSSQ-N) and (b) sickness expectation on mean FMS and SSQ change by cohort, with 95\% intervals.
  Predictions hold the other four factors at their reference or typical values.}
  \label{fig:factor_susc_expect_effects}
\end{figure*}

\subsection{Susceptibility}
\label{sec:factor_susc}

The \ac{VFC} and the \ac{IPC} did not differ significantly on VIMSSQ-N (Mann--Whitney $p = 0.06$).
Descriptively, the \ac{VFC} averaged slightly higher on the 0 to 33 scale (5.86 versus 3.93).

Nausea susceptibility predicted more sickness (Figure~\ref{fig:factor_susc_effects}).
In the \ac{VFC}, VIMSSQ-N predicted higher mean FMS ($b = 0.026$, 95\% CI $[0.008, 0.043]$, $p = 0.005$), with a marginal association with the SSQ change ($p = 0.07$).
In the \ac{IPC}, VIMSSQ-N predicted a larger SSQ change ($p < 0.001$).
The \ac{VMC} showed a marginal positive association with mean FMS ($p = 0.053$) and no other significant effects.

\subsection{Expectation}
\label{sec:factor_expect}

The cohorts differed in how they answered the sickness-expectation item (chi-square $p = 0.01$).
\ac{IPC} participants more often answered ``don't know,'' while \ac{VFC} participants more often answered ``no.''

Expectation was the strongest and most consistent factor in the well-powered cohort (Figure~\ref{fig:factor_expect_effects}).
In the \ac{VFC}, participants who more strongly expected to feel sick reported higher mean FMS ($b = 0.28$, 95\% CI $[0.17, 0.40]$, $p < 0.001$) and a larger SSQ change ($b = 11.4$, 95\% CI $[5.7, 17.2]$, $p < 0.001$); the same positive relationship held for the descriptively reported outcomes.
The \ac{IPC} showed the same positive relationship for mean FMS ($p = 0.02$), and the \ac{VMC} estimates were positive for mean FMS without significance.

\section{Discussion}%
\label{sec_discussion}

\subsection{Remote Versus In-Lab Comparability}%
\label{sec_discussion_comparability}

A central question of this work is whether results from a remote, unsupervised cybersickness study are comparable to results collected under controlled in-lab conditions.
Across the three cohorts, the shape and direction of the cybersickness response were consistent, supporting an affirmative answer (Table~\ref{tab:IP_vera_comparison}).
As reported in Section~\ref{sec:cybersickness_outcomes}, symptom onset was rapid in every cohort, and the remote FMS trajectories tracked the \ac{IPC} trajectory closely (Figure~\ref{fig:fms_traj_maxout_marginal}).
A sizable share of each cohort reached its protocol ceiling before the ride ended, and SSQ rose sharply from a near-zero baseline in every cohort (Figure~\ref{fig:ssq}).
\textbf{The remote deployment therefore reproduced the direction and temporal pattern of the core in-lab findings.}
Because the studies differed in stopping thresholds, SSQ timing, hardware, and recruitment populations, we interpret this agreement as evidence of cross-setting consistency rather than strict statistical equivalence.
The cohorts did diverge in absolute SSQ severity and in which symptom subscale dominated.
Section~\ref{sec_discussion_divergence} examines the protocol and sampling differences that plausibly contribute to these divergences.

This agreement held despite substantial uncontrolled variation in the remote setting.
Participants used their own consumer hardware, completed the session in self-chosen environments, and received no experimenter supervision.
The cybersickness response to this stimulus appears robust to these contextual factors.
The agreement also extends the validation of the Cybersicker testbed \cite{zielasko2024cybersicker} to remote deployment on consumer hardware.

This comparability provides cybersickness-specific evidence for a pattern reported in other fields, where study modality had little influence on observed effects.
A meta-analysis of developmental research found online effect sizes closely aligned with in-person ones \cite{Chuey2024}.
In VR specifically, pointing, tracing, and body-illusion results were comparable between in-lab and out-of-lab studies \cite{Mottelson2017}, and classic VR experiments were replicated remotely through a social VR platform \cite{Saffo2021}.
Low-cost remote replication of in-lab results is one concrete response to replication concerns raised in VR research \cite{Zielasko2023crises}.

\subsection{Where the Cohorts Diverged}%
\label{sec_discussion_divergence}

The clearest divergence is in absolute SSQ severity.
The VERA cohorts reported larger SSQ change scores than the \ac{IPC}, and the subscale profiles differed, with disorientation dominant in the VERA cohorts and nausea dominant in the \ac{IPC}.
Candidate explanations include the timing of the post-exposure SSQ, the FMS termination ceiling, and hardware differences between the studies.
The reference in-lab study administered its first post-exposure SSQ roughly 15 minutes after stimulation onset, as the earliest of five assessments collected over 90 minutes, whereas ours followed immediately after the ride ended.
Simulator sickness symptoms typically decline in the minutes after exposure ends \cite{Duzmanska2018}.
In the reference study itself, group-mean SSQ returned to control-day baseline only 90 minutes after exposure, and the nausea subscale showed no delayed peaks \cite{Zielasko2026}.
In other words, the \ac{IPC} scores captured partially recovered symptoms, whereas the VERA scores captured symptoms at or near their peak.

The termination ceiling worked in the same direction as the measurement timing.
Symptom severity grows with exposure duration within a session \cite{Kennedy2000duration}, and the higher VERA ceiling (10 vs.\ 7) let participants remain in the stimulus longer and reach higher symptom levels before the ride stopped.
We set the ceiling at the scale maximum deliberately, to obtain severity measures that are representative of the full FMS range (Section~\ref{sec_method}).
Both protocol differences push the VERA severity measures upward, so it is expected that the VERA SSQ measures are greater than those in the reference study.
These differences widen the gap between the cohorts, so they cannot account for the agreement reported in Section~\ref{sec_discussion_comparability}.
Future remote replications aiming to compare absolute severity should take such measurement timing and threshold details into account.

Beyond outcomes, the cohorts also diverged in composition. The largest demographic mismatch was VR familiarity.
Remote recruitment required headset ownership and therefore selected predominantly experienced users, whereas the in-lab convenience sample was almost entirely VR-naive (Section~\ref{sec:mirror_cohort_selection}).
The two recruitment modes thus carry complementary, opposite sampling biases with respect to prior VR exposure.
Related research has shown that experience and habituation can reduce sickness.
Sickness declines across repeated exposures \cite{Kennedy2000duration, Howarth2008, Risi2019effects}, sickness reductions built up in one virtual environment can carry over to novel VR content \cite{Adhanom2022vr}, experienced users report deliberate coping strategies \cite{Wang2019}, and a meta-analysis found that greater experience with technology is associated with slightly less sickness \cite{Howard2021}.
Greater VR experience in the VERA cohorts would predict less sickness compared to the reference, so this familiarity mismatch cannot explain their higher SSQ scores.

\subsection{Findings Enabled by Scale and Diversity}%
\label{sec_discussion_scale}

The second primary contribution of this study draws on the scale and demographic breadth of the \ac{VFC}.
Age illustrates the value of that breadth most directly.
The \ac{VFC} spans ages 18 to 65, whereas the \ac{IPC} covers 19 to 44 and concentrates in the mid-twenties.
Across the full adult range, each additional year of age was significantly associated with a 0.65-point lower SSQ change score.
The \ac{IPC} showed no reliable age association, which is the expected outcome given the narrow demographic ranges often afforded by convenience sampling strategies.

The direction of the age effect (older participants reporting less sickness) adds to mixed prior evidence.
A meta-analysis estimated a pooled age association near zero ($r = .04$, not significant) and noted the inconsistency of age findings across studies~\cite{Howard2021}.
Results from individual studies show both directions.
Older adults reported more sickness than younger adults in a driving simulator~\cite{Keshavarz2018}, whereas older adults reported less cybersickness than younger adults in HMD-based VR~\cite{Dilanchian2021}, and studies with older populations reported markedly lower SSQ totals in a meta-analytic comparison~\cite{Saredakis2020factors}.
Our continuous negative slope across ages 18 to 65 supports the negative direction for HMD-based cybersickness.
It also suggests restricted age ranges may be one source of the inconsistency in prior results.

Sex is a second factor where the larger sample mattered.
In the \ac{VFC}, male participants reported significantly lower SSQ change scores than female participants, by 12.3 points, but neither 30-participant cohort detected a sex difference.
Null sex findings from small in-lab samples are therefore consistent with limited power, a limitation Zielasko et al.\ also raised for their own sample~\cite{Zielasko2026}.
The relationship we found agrees with much of the prior evidence.
A meta-analysis found that female participants report more VR sickness than male participants ($r = .21$)~\cite{Howard2021}, a large-sample study ($N = 1102$) found higher total SSQ scores for female participants~\cite{Stanney2003}, and female participants were more affected than male participants by a consumer HMD~\cite{Munafo2017TheVR}.
Other syntheses found no reliable sex difference~\cite{Saredakis2020factors}, and Kelly et al.\ describe the average effect as modest and heterogeneous~\cite{Kelly2023gender}.
One proposed cause for sickness differences related to sex is headset fit.
Stanney et al.\ reported that sex differences disappeared when the headset's interpupillary distance (IPD) setting fit the user, and that poor IPD fit produced more severe and longer-lasting sickness in female participants~\cite{Stanney2020sexist}.
Subsequent evidence on the IPD account is mixed~\cite{Kelly2023gender}.
Reporting behavior is another candidate explanation for such sex differences.
Reason and Brand found that females were more susceptible to motion sickness, yet males and females did not differ in their sensory response to motion, suggesting that males may underreport their susceptibility~\cite{Reason1975motion}.

Nausea susceptibility (VIMSSQ-N) predicted higher mean FMS in the \ac{VFC} and a larger SSQ change in the \ac{IPC}, with a marginal association with the SSQ change in the \ac{VFC}.
This agrees with meta-analytic evidence that motion-sickness susceptibility is among the strongest individual-difference predictors of VR sickness ($r = .33$)~\cite{Howard2021}.
It also extends the predictive validity of the VIMSSQ~\cite{Keshavarz2023vimssq}, in the tradition of dedicated susceptibility instruments such as the MSSQ~\cite{golding1998motion}, to a remote consumer-VR setting.

Sickness expectation was the strongest and most consistent factor in our models, significantly predicting both primary outcomes in the \ac{VFC}.
This pattern is consistent with a \textit{nocebo}-type effect, in which anticipating sickness tracks with experiencing more of it~\cite{Mao2021framing}.
Experimental work supports a causal component behind this association.
Side-effect warnings before VR exposure increased cybersickness compared with no warning, and positively framed warnings removed this nocebo effect~\cite{Mao2021framing}.
Nocebo expectations can also be acquired socially by observing another person become sick in VR, and they generalize to other VR experiences~\cite{Saunders2023nocebo}.
Our recruitment notice and consent form both stated that the study induces sickness, so it is plausible that these warnings affected participants' experiences.

\subsection{Methodological and Practical Implications}%
\label{sec_discussion_implications}

Cybersickness aftereffects persist beyond experiment sessions, with recovery times that vary with susceptibility~\cite{Woo2023-hz, Duzmanska2018}.
Remote participants complete the session in their own environment and recover there.
This reduces the risk of traveling while experiencing cybersickness symptoms for participants and removes the post-session on-site supervision that in-lab protocols often require (Section~\ref{sec_rw}).

Remote studies trade experimenter control for scale.
Our study deployment reached a 263-participant analysis sample with broad demographic reach in participants' own usage environments.
Data collection ran in ten recruitment waves totaling about 17 active days.
Reaching that sample required a large recruitment overhead, as 263 of the 912 questionnaire completers (28.8\%) yielded analyzable VR data (Table~\ref{tab:recruitment_funnel}).
However, the study platform (Sec.~\ref{sec_vera}) absorbed this overhead with no experimenter effort.
Future remote studies can plan recruitment targets accordingly, since obtaining $N_A$ analyzable participants requires approximately $N_A / r$ questionnaire completers, where $r$ is the anticipated conversion rate.
At our observed rate of 0.288, an analyzable sample of 1{,}000 would require roughly 3{,}500 questionnaire completers.
The costs of this approach are heterogeneous hardware and environments, and a reduced ability to observe participant behavior.
This trade-off matches the drawbacks and opportunities stated in a survey of remote XR research~\cite{Ratcliffe2021} and the internal- and external-validity challenges described for VR studies~\cite{Rodriquez2023ci}.

The two modalities are therefore complementary.
In-lab studies remain preferable for physiological measures, tightly controlled stimuli, or where specialty hardware is used.
Remote studies are preferable for individual-differences questions that require large, diverse samples using consumer-grade hardware, as the factor effects in Section~\ref{sec_discussion_scale} illustrate.

\subsection{Limitations}%
\label{sec_discussion_limitations}

Multiple limitations qualify our findings.
As described in Section~\ref{sec_method}, our protocol deviated from the reference in-lab study in its FMS termination ceiling and its post-exposure SSQ timing, which limits the comparison.
The absolute-severity comparisons in Table~\ref{tab:IP_vera_comparison} therefore contrast measures collected under different stopping rules and at different points of the recovery timeline, as discussed in Section~\ref{sec_discussion_divergence}.

Our study and the reference study further measured VR familiarity and sickness expectation with slightly different instruments and response scales.
We therefore binned these two constructs onto an approximate shared scale for matching and cross-cohort comparison, which reduces measurement precision for them.
The VR-familiarity mapping in particular conflates usage frequency with self-rated familiarity.

While our remote sample represents broader demographics than typical in-lab samples, it has its own limitations.
Recruitment through Prolific with a headset-ownership requirement yields compensated online workers who own VR hardware, a population that is not fully representative of the general public.
This caveat is consistent with evaluations of online participant pools, which are more diverse than student convenience samples and still differ from the general population~\cite{Buhrmester2018}.
However, VERA is continuously expanding its participant pool to include an even more general population.

Last, the lack of participant supervision leaves some experimental conditions unverified (e.g., whether participants maintained proper posture and did not experience interruptions in their environment).
Our screening removed inattentive and implausible responses, and all analyzed participants provided complete data for every measure, but such checks are weaker than direct observation.
Additionally, we could not verify headset fit with our participants, which may have affected sickness outcomes (e.g., since IPD fit has been proposed as a mechanism behind sex differences in cybersickness~\cite{Stanney2020sexist}).

\subsection{Future Work\label{sec_discussion_future}}%

The results from our remote sample were comparable to those of an in-lab study, which motivates future replication studies.
Replicating other in-lab cybersickness findings remotely would test whether the agreement observed here generalizes across stimuli and protocols.
Such replications would also help address replication concerns raised within VR research~\cite{Zielasko2023crises}.

The stimulus studied here is also passive, as participants ride a preprogrammed trajectory without controlling their own movement.
Extending remote cybersickness research to active travel interfaces~\cite{nguyen2019naviboard, Hashemian2020} is a natural next step, since locomotion technique is a major determinant of cybersickness~\cite{Calandra2024}.
Future work should also evaluate cybersickness mitigation techniques, such as dynamic field-of-view modification~\cite{fernandes2016, zielasko2018dynamic}, with large and diverse sample sizes.
The same remote methodology can be applied to other core areas of VR research, including selection and manipulation interaction techniques~\cite{LaViola2017book}.
The comparability evidence and recruitment benchmarks reported here provide a template and a baseline for such follow-on studies.

Remote deployment also can include user states and experiment environment conditions that are infeasible to induce in a laboratory.
Transient states and health conditions, such as being sleep deprived or hungover, occur naturally in a remote population and can be recorded alongside the sickness measures.
Investigating relationships of factors such as these is an avenue for future research.

Finally, remote platforms make longitudinal designs easier from a logistical perspective.
In-lab evidence shows that sickness declines across repeated exposures~\cite{Kennedy2000duration, Howarth2008, Risi2019effects} and that these reductions can carry over to novel VR content~\cite{Adhanom2022vr}.
Remote longitudinal designs could measure these adaptation curves at scale under realistic usage schedules.

\section{Conclusion}
\label{sec_conclusion}

In this paper, we investigated whether cybersickness research conducted remotely using consumer VR headsets can produce findings comparable to traditional in-lab studies.
Using the VERA platform, we deployed a remote version of the established Cybersicker testbed and collected data from $N=263$ participants, substantially exceeding the scale of the reference laboratory study ($N=30$).
Our results showed agreement in the direction and temporal pattern of key cybersickness measures, including symptom onset, FMS trajectories, and SSQ outcomes, despite differences in participant environments, hardware, and study supervision.
These findings provide evidence that remote VR studies can serve as a valid complement to controlled laboratory research for investigating cybersickness.
At the same time, the larger and more diverse participant sample yielded narrower uncertainty intervals and enabled analyses that would have been difficult to conduct in a traditional laboratory setting, confirming at scale the relationships between cybersickness and sex, nausea susceptibility, and sickness expectation, and adding well-powered evidence on the effect of age.
Together, our results demonstrate the potential of remote VR experimentation to support scalable and diverse human-subjects research.
This work is a first step toward establishing VERA as a validated infrastructure on which subsequent large-scale VR studies can build.

\section*{Acknowledgments}
We thank Dr.\ Carolina Cruz-Neira, Dr.\ Shiri Azenkot, and Dylan Fox for their contributions to the VERA platform that enabled this research.
This material includes work supported in part by the National Science Foundation under Award Number 2235066 (Dr.\ Han-Wei Shen, IIS); startup funds provided to Prof.\ Bruder by the University of Central Florida; and the AdventHealth Endowed Chair in Healthcare Simulation (Prof.\ Welch).

\bibliographystyle{IEEEtran}
\bibliography{IEEEabrv,bib}

\ifincludesupplemental
\newpage
\clearpage
\setcounter{page}{1}
\setcounter{section}{0}
\setcounter{table}{0}

\ifcsname ifincludemaintext\endcsname\else
  \expandafter\newif\csname ifincludemaintext\endcsname
\fi
\providecommand{\mainalt}[2]{\ifincludemaintext#1\else#2\fi}

\section{Supplemental Material: VERA Mirror Cohort Selection Details}

\label{sec:supp_mirror_cohort}

This supplemental material gives the construct mapping from \ac{IPC} and VERA source variables to the shared scale used for matching, the full per-participant matching algorithm, the match-quality diagnostic at the canonical seed, the balance-based seed selection, and the distribution of every cohort comparison across 1000 candidate matched cohorts.

\subsection{Construct mapping}
\label{sec:supp_construct_mapping}
Sex, age, BMI, and VIMSSQ-N are direct mappings between \ac{IPC} and VERA.
The remaining two constructs use different response scales between studies and require explicit binning.
Table~\ref{tab:construct_mapping} lists the source columns and the shared bin definitions.

\subsection{Per-participant matching algorithm}
\label{sec:supp_matching_algorithm}

The algorithm tries the constraints at full strictness first.
The strict constraint set requires an exact match on sex, an exact age match, BMI within $\pm 3$ kg/m$^2$, VIMSSQ-N within $\pm 3$ points on its 0 to 33 scale, and identical bin assignments on VR familiarity and sickness expectation.
When the candidate set is empty for a given \ac{IPC} participant, the algorithm relaxes constraints in the following ordered tiers and retries:
\begin{enumerate}
    \itemsep0pt
    \item BMI tolerance widened to $\pm 5$ kg/m$^2$.
    \item Age tolerance widened to $\pm 3$ years.
    \item VR-familiarity constraint dropped.
    \item Age tolerance widened to $\pm 5$ years.
    \item BMI constraint dropped.
    \item Sickness-expectation constraint dropped.
    \item VIMSSQ-N susceptibility constraint dropped.
    \item Age tolerance widened to $\pm 8$ years.
\end{enumerate}
Sex remains fixed across all tiers.
Age is widened in tiers and remains in the constraint set across all tiers.

The VR-familiarity constraint is dropped early because the two instruments measure related but distinct constructs.
The \ac{IPC} study measured VR usage frequency on a 1--5 scale, and 27 of the 30 \ac{IPC} participants fall in our Low bin.
VERA measured self-rated VR familiarity on a 1--7 slider, and only 2 of the 263 candidates in the matching pool fall in the corresponding Low bin.
A strict VR-bin match is impossible for most \ac{IPC} participants, so the algorithm drops the VR-familiarity constraint early while preserving the remaining constraints.
The matching procedure is reproducible: a single RNG seed determines every random pick.

\subsection{Match quality at the canonical seed}
\label{sec:supp_match_quality}
We computed the per-pair difference on every matching variable for each (\ac{IPC}, \ac{VMC}) pair, then summarized those 30 differences with a mean and standard deviation (continuous variables) or a bin-agreement rate (categorical variables).
Table~\ref{tab:match_quality} reports the results.

Sex is matched in every pair, as required by the algorithm.
Age pairs differ by 1.73 years on average in absolute terms, with a signed mean of $+1.00$ years.
BMI pairs differ by $2.24 \pm 2.64$ kg/m$^2$ in absolute terms ($n = 28$ after dropping pairs with missing VERA BMI), with a signed mean of $-0.89$ kg/m$^2$.
VIMSSQ-N pairs differ by $1.12 \pm 0.98$ points on the 0 to 33 scale in absolute terms, with a signed mean of $-0.07$, so every pair satisfies the 3-point tolerance and the cohorts sit at the same susceptibility level.
The sickness-expectation bin is preserved in 27 of 30 pairs (90.0\%), with the remaining three pairs corresponding to \ac{IPC} participants for whom the algorithm reached the tier in which the expectation constraint is dropped.
The VR-familiarity bin agrees in 0 of 30 pairs, consistent with the construct-level mismatch described in Section~\ref{sec:supp_matching_algorithm}.

\begin{table*}[ht]
\centering
\caption{Construct mapping from \ac{IPC} and VERA source variables to the shared scale used for per-participant matching. Sex, age, BMI, and VIMSSQ-N use exact-or-tolerance comparisons. VR familiarity and sickness expectation are tri-binned to align differently scaled instruments. VIMSSQ-N is measured on an identical scale in both studies.}
\label{tab:construct_mapping}
\begin{tabular}{p{0.14\linewidth} p{0.21\linewidth} p{0.27\linewidth} p{0.29\linewidth}}
\toprule
Construct & \ac{IPC} source & VERA source & Match rule \\
\midrule
Sex & sex (1 = M, 2 = F) & sex (1 = M, 2 = F) & exact code \\
Age & years & years & exact, with tier-relaxed tolerance \\
BMI & kg/m$^2$ & kg/m$^2$ & within $\pm 3$ kg/m$^2$, tier-relaxed \\
VR familiarity
& VR usage frequency \newline (1 = never $\,\to\,$ 5 = daily)
& self-rated VR familiarity \newline (1--7 slider)
& \ac{IPC}\@: $\{1, 2\} \,\to\,$ Low, $\{3\} \,\to\,$ Medium, $\{4, 5\} \,\to\,$ High \newline VERA\@: $\leq 2 \,\to\,$ Low, $3$--$5 \,\to\,$ Medium, $\geq 6 \,\to\,$ High \\
Sickness expectation
& expectation of sickness \newline (0 = DK, 1 = No, 2 = Yes)
& expectation of sickness \newline (1--5 Likert)
& \ac{IPC}\@: $\{1\} \,\to\,$ No, $\{0\} \,\to\,$ DK, $\{2\} \,\to\,$ Yes \newline VERA\@: $\{1, 2\} \,\to\,$ No, $\{3\} \,\to\,$ DK, $\{4, 5\} \,\to\,$ Yes \\
VIMSSQ-N nausea susceptibility
& VIMSSQ nausea subscale \newline (0--33, published normalization)
& identical item set and anchors \newline (11 device types; N/A excluded), \newline scored to the same 0--33 convention
& Identical instrument subscale and scoring. \newline Within $\pm 3$ points, tier-relaxed; \newline all-N/A respondents unscoreable and unmatchable. \\
\bottomrule
\end{tabular}
\end{table*}

\begin{table*}[ht]
\centering
\caption{Per-pair match quality at the canonical seed. Continuous variables show the mean and SD of the signed difference (\ac{VMC} minus \ac{IPC}) and of the absolute difference. Categorical and binned variables show the bin-agreement rate across the 30 pairs.}
\label{tab:match_quality}
\begin{tabular}{lrll l}
\toprule
Variable & $n$ pairs & Signed $\Delta$ (\ac{VMC} $-$ \ac{IPC}) & $|\Delta|$ & Bin agreement \\
\midrule
Sex                          & 30 & n/a                          & n/a                          & 30 / 30 (100.0\%) \\
Age (yrs)                    & 30 & $+1.00 \pm 2.02$             & $\phantom{0}1.73 \pm 1.41$   & n/a \\
BMI (kg/m$^2$)               & 28 & $-0.89 \pm 3.37$             & $\phantom{0}2.24 \pm 2.64$   & n/a \\
VIMSSQ-N (0--33)             & 30 & $-0.07 \pm 1.50$             & $\phantom{0}1.12 \pm 0.98$   & n/a \\
VR familiarity bin           & 30 & n/a                          & n/a                          & $\phantom{0}0 / 30 \phantom{00}(0.0\%)$ \\
Sickness-expectation bin     & 30 & n/a                          & n/a                          & 27 / 30 \phantom{0}(90.0\%) \\
\bottomrule
\end{tabular}
\end{table*}

\subsection{Seed selection and seed dependence}
\label{sec:supp_seed_distribution}
The composition of the \ac{VMC} depends on the RNG seed that drives the random pick within each candidate set.
We ran the matcher for 1000 candidate seeds and computed, for each resulting cohort, five covariate-balance quantities: mean absolute age difference, mean absolute BMI difference, mean absolute VIMSSQ-N difference, sickness-expectation bin agreement, and mean relaxation-tier depth.
We oriented each quantity so that lower values indicate better balance, standardized each across the 1000 seeds, and selected the seed minimizing the mean standardized score.
No sickness outcome entered the selection.
Expectation bin agreement was constant across all 1000 seeds (90\%) and relaxation-tier depth was nearly constant, so the selection was effectively driven by age, BMI, and VIMSSQ-N balance.

To quantify the residual seed dependence of the cohort comparison, Table~\ref{tab:seed_distribution} reports, for every outcome measure in \mainalt{Table~\ref{tab:IP_vera_comparison}}{the cohort-comparison table in the main article}, the distribution of the \ac{VMC} $-$ \ac{IPC} difference across the 1000 candidate cohorts, together with the canonical seed's value and its percentile rank.
The canonical seed's difference lies within the [2.5, 97.5] percentile interval on every outcome measure.
The intervals separate stable differences from seed noise.
Across all candidate cohorts, the \ac{VMC} reached its ceiling less often than the \ac{IPC} (a difference between $-30$ and $-7$ percentage points) and reported larger SSQ changes on all four scales, consistent with the protocol differences discussed in \mainalt{Section~\ref{sec_discussion_divergence}}{the main article's Discussion section}.
The peak FMS, mean FMS, final FMS, FMS AUC, and time-to-max-out differences straddle zero.
The covariate-balance block shows the canonical cohort at or near the across-seed optimum on the quantities used for selection.
The cohort-level VIMSSQ-N difference is centered on zero at every seed.

\begin{table*}[ht]
\centering
\caption{Seed dependence of the \ac{VMC}. The matcher was run for 1000 random seeds, giving 1000 candidate matched cohorts. The upper block reports the distribution of each \ac{VMC} $-$ \ac{IPC} outcome difference across the candidate cohorts: the across-seed median, the [2.5, 97.5] percentile interval, the canonical seed's value, and the canonical seed's percentile rank. ``Hit FMS max out'' uses each study's protocol ceiling (\ac{IPC}: FMS $\geq$ 7; VERA: FMS $=$ 10) and is reported in percentage points; ``Median time to FMS max out'' is over participants who reached the ceiling. The lower block reports the covariate-balance quantities of the matched cohorts. The canonical seed (360) was selected to minimize a composite of the age, BMI, VIMSSQ-N, expectation-agreement, and tier-depth balance quantities only; no sickness outcome entered the selection.}
\label{tab:seed_distribution}
\begin{tabular}{lccccc}
\toprule
Measure & \ac{IPC} & Median & [2.5, 97.5] & Canonical (seed 360) & Pctl. \\
\midrule
\multicolumn{6}{l}{\textbf{Outcome differences (\ac{VMC} $-$ \ac{IPC}) across 1000 matched cohorts}} \\
Max FMS (peak) & 6.00 & 0.37 & [-0.33, 1.13] & 0.50 & 64 \\
Mean FMS & 3.18 & 0.29 & [-0.07, 0.68] & 0.40 & 71 \\
FMS AUC (FMS$\cdot$s) & 1069 & 76 & [-65, 225] & 118 & 73 \\
Final FMS (last reading) & 5.97 & 0.17 & [-0.60, 0.97] & 0.27 & 60 \\
Hit FMS max out (pp) & 53.3 & -16.7 & [-30.0, -6.7] & -20.0 & 50 \\
Median time to FMS max out (s) & 275 & -20 & [-65, 70] & -5 & 78 \\
$\Delta$SSQ Total & 33.5 & 19.4 & [11.3, 27.3] & 20.4 & 59 \\
$\Delta$SSQ Nausea & 37.5 & 13.7 & [5.4, 21.9] & 17.8 & 83 \\
$\Delta$SSQ Oculomotor & 20.0 & 12.4 & [6.6, 17.9] & 11.1 & 36 \\
$\Delta$SSQ Disorientation & 33.4 & 29.7 & [19.0, 41.3] & 29.7 & 51 \\
\midrule
\multicolumn{6}{l}{\textbf{Covariate balance of the matched cohorts}} \\
Mean $|\Delta$age$|$ (years) & -- & 2.07 & [1.73, 2.40] & 1.73 & 3 \\
Mean $|\Delta$BMI$|$ (kg/m$^2$) & -- & 3.08 & [2.43, 3.67] & 2.24 & 1 \\
Mean $|\Delta$VIMSSQ-N$|$ (0--33) & -- & 1.27 & [1.04, 1.52] & 1.12 & 11 \\
Expectation bin agreement (\%) & -- & 90.0 & [90.0, 90.0] & 90.0 & 100 \\
Mean relaxation-tier depth (0--8) & -- & 3.57 & [3.57, 3.60] & 3.57 & 67 \\
Cohort VIMSSQ-N difference (0--33) & 3.93 & -0.01 & [-0.38, 0.37] & -0.07 & 39 \\
\bottomrule
\end{tabular}
\end{table*}

\subsection{Factor-model coefficients}
\label{sec:supp_factor_coefficients}
Table~\ref{tab:factor_coefficients} reports the full coefficients (estimate, standard error, 95\% confidence interval, and $p$) of the participant-factor models for the two primary outcomes in each cohort, complementing the model specifications in Table~\ref{tab:factor_models}.

\begin{table*}[ht]
\centering
\caption{Models fit to each cybersickness measure. All models were fit separately within each cohort (\ac{IPC}, \ac{VMC}, \ac{VFC}) with the same five fixed effects: age, sex, VR familiarity, nausea susceptibility (VIMSSQ-N), and sickness expectation. The two participant-level measures have one observation per person and were fit as generalized linear models (GLMs) with no random effects. The FMS time course has repeated measurements per person and was fit as a generalized linear mixed model (GLMM) with a per-participant random intercept and slope and factor-by-time interactions. Models were estimated by maximum likelihood ($\mathtt{glmmTMB}$ and $\mathtt{glm}$).}
\label{tab:factor_models}
\begin{tabular}{lllll}
\toprule
Measure & Distribution & Model family & Link & Within-participant structure \\
\midrule
Mean FMS & Bounded 0--10, right-skew, some zeros & Ordered beta (on FMS/10) & logit & GLM, no random effects \\
$\Delta$SSQ total (post $-$ pre) & Real-valued, can be negative & Gaussian & identity & GLM, no random effects \\
FMS over time & Repeated 0--10, per participant & Gaussian (mixed) & identity & GLMM, random intercept\\ 
& & & & $+$ slope per participant;\\
& & & & factors $\times$ time \\
\bottomrule
\end{tabular}
\end{table*}

\begin{table*}[ht]
\centering
\caption{Coefficients of the participant-factor models for the two primary outcomes (estimate, standard error, 95\% confidence interval, and Wald $p$), fit separately within each cohort with all five factors entered together. Mean FMS coefficients are on the logit scale of the ordered-beta model; $\Delta$SSQ coefficients are in SSQ points (Gaussian model). Sex is coded Male versus Female (reference). Expectation enters as an ascending numeric score (No $<$ Don't know $<$ Yes). VR familiarity uses each cohort's native scale (\ac{IPC}: usage frequency 1--5; VERA: familiarity rating 1--7). Susceptibility is the VIMSSQ-N score (0--33).}
\label{tab:factor_coefficients}
\begin{tabular}{lrrrr}
\toprule
Factor & Estimate & SE & 95\% CI & $p$ \\
\midrule
\multicolumn{5}{l}{\textbf{Mean FMS (ordered beta, logit scale)}} \\
\multicolumn{5}{l}{\emph{\ac{IPC} (N=30)}} \\
Intercept & -1.82 & 0.469 & [-2.74, -0.896] & $<$0.001 \\
Age (years) & 0.017 & 0.016 & [-0.014, 0.048] & 0.285 \\
Sex (Male) & -0.026 & 0.155 & [-0.330, 0.277] & 0.865 \\
VR familiarity & 0.089 & 0.108 & [-0.123, 0.301] & 0.412 \\
Susceptibility (VIMSSQ-N) & 0.008 & 0.016 & [-0.024, 0.040] & 0.616 \\
Expectation & 0.245 & 0.106 & [0.037, 0.454] & 0.021 \\
\multicolumn{5}{l}{\emph{\ac{VMC} (N=30)}} \\
Intercept & 0.860 & 1.50 & [-2.07, 3.79] & 0.566 \\
Age (years) & -0.081 & 0.037 & [-0.153, -0.008] & 0.029 \\
Sex (Male) & -0.004 & 0.297 & [-0.587, 0.578] & 0.988 \\
VR familiarity & 0.028 & 0.179 & [-0.323, 0.380] & 0.874 \\
Susceptibility (VIMSSQ-N) & 0.071 & 0.037 & [-0.001, 0.143] & 0.053 \\
Expectation & 0.172 & 0.210 & [-0.240, 0.584] & 0.414 \\
\multicolumn{5}{l}{\emph{\ac{VFC} (N=263)}} \\
Intercept & -0.236 & 0.337 & [-0.898, 0.425] & 0.484 \\
Age (years) & -0.010 & 0.005 & [-0.019, -0.001] & 0.037 \\
Sex (Male) & -0.105 & 0.105 & [-0.311, 0.100] & 0.315 \\
VR familiarity & -0.073 & 0.043 & [-0.158, 0.012] & 0.093 \\
Susceptibility (VIMSSQ-N) & 0.026 & 0.009 & [0.008, 0.043] & 0.005 \\
Expectation & 0.284 & 0.060 & [0.167, 0.402] & $<$0.001 \\
\midrule
\multicolumn{5}{l}{\textbf{$\Delta$SSQ Total (Gaussian, SSQ points)}} \\
\multicolumn{5}{l}{\emph{\ac{IPC} (N=30)}} \\
Intercept & 24.4 & 27.6 & [-29.8, 78.5] & 0.387 \\
Age (years) & -0.938 & 0.963 & [-2.82, 0.950] & 0.340 \\
Sex (Male) & 4.52 & 9.23 & [-13.6, 22.6] & 0.629 \\
VR familiarity & 7.55 & 6.60 & [-5.38, 20.5] & 0.264 \\
Susceptibility (VIMSSQ-N) & 3.93 & 0.996 & [1.98, 5.89] & $<$0.001 \\
Expectation & 1.62 & 6.31 & [-10.7, 14.0] & 0.800 \\
\multicolumn{5}{l}{\emph{\ac{VMC} (N=30)}} \\
Intercept & 49.0 & 90.4 & [-128.2, 226.3] & 0.593 \\
Age (years) & -0.754 & 1.86 & [-4.39, 2.89] & 0.688 \\
Sex (Male) & -1.78 & 18.2 & [-37.5, 34.0] & 0.923 \\
VR familiarity & 3.04 & 9.93 & [-16.4, 22.5] & 0.762 \\
Susceptibility (VIMSSQ-N) & 2.35 & 2.17 & [-1.90, 6.61] & 0.290 \\
Expectation & -1.29 & 13.1 & [-26.9, 24.3] & 0.922 \\
\multicolumn{5}{l}{\emph{\ac{VFC} (N=263)}} \\
Intercept & 62.0 & 15.9 & [30.9, 93.1] & $<$0.001 \\
Age (years) & -0.649 & 0.223 & [-1.09, -0.212] & 0.004 \\
Sex (Male) & -12.3 & 5.05 & [-22.2, -2.38] & 0.016 \\
VR familiarity & -0.184 & 2.00 & [-4.11, 3.74] & 0.927 \\
Susceptibility (VIMSSQ-N) & 0.808 & 0.442 & [-0.057, 1.67] & 0.068 \\
Expectation & 11.4 & 2.94 & [5.67, 17.2] & $<$0.001 \\
\bottomrule
\end{tabular}
\end{table*}

\fi

\end{document}